\documentclass[11pt]{article}
\usepackage[margin=1in]{geometry}
\usepackage{booktabs} 
\usepackage{authblk}
\usepackage{amsmath, amssymb, bm, amsthm, bbm} 
\usepackage[ruled]{algorithm2e} 

\newtheorem{assumption}{Assumption}
 
\newtheorem{proposition}{Proposition}
\theoremstyle{definition}
\newtheorem{definition}{Definition}

\newtheorem{example}{Example}
\usepackage{bm, bbm} 
\usepackage{hyperref} 
\usepackage{subcaption}
\usepackage{natbib} 
\usepackage{hyperref}
\usepackage{tikz} 
\usepackage{setspace}
\setcitestyle{authoryear}

\usetikzlibrary{shapes,decorations,arrows,calc,arrows.meta,fit,positioning}
\tikzset{
    -Latex,auto,node distance =1 cm and 1 cm,semithick,
    state/.style ={ellipse, draw, minimum width = 0.7 cm},
    point/.style = {circle, draw, inner sep=0.04cm,fill,node contents={}},
    bidirected/.style={Latex-Latex,dashed},
    el/.style = {inner sep=2pt, align=left, sloped}
}

\begin{document}

\title{Causal Estimation of User Learning in Personalized Systems\thanks{We are grateful to Steve Howard, participants at the 2022 Conference on Digital Experimentation @ MIT (CODE@MIT), as well as the anonymous referees for EC'23 for helpful comments and suggestions.}}
\author[1]{Evan Munro}
\author[2]{David Jones\footnote{These authors contributed equally to this work.}}
\author[2]{Jennifer Brennan$^\dagger$}
\author[2]{Roland Nelet}
\author[2]{Vahab Mirrokni}
\author[2]{Jean Pouget-Abadie}
\affil[1]{Stanford Graduate School of Business}
\affil[2]{Google Research}
\maketitle

\begin{abstract}

In online platforms, the impact of a treatment on an observed outcome may change over time as 1) users learn about the intervention, and 2) the system personalization, such as individualized recommendations, change over time. We introduce a non-parametric causal model of user actions in a personalized system. We show that the Cookie-Cookie-Day (CCD) experiment, designed for the measurement of the user learning effect, is biased when there is personalization. We derive new experimental designs that intervene in the personalization system to generate the variation necessary to separately identify the causal effect mediated through user learning and personalization. Making parametric assumptions allows for the estimation of long-term causal effects based on medium-term experiments. In simulations, we  show that our new designs successfully recover the dynamic causal effects of interest.  
\end{abstract} 

\newpage

\section{Introduction}\label{sec:intro}


In many online platforms, the immediate impact of a treatment, such as a UI change or algorithm change, is not equal to its long-term impact due to a variety of indirect effects that occur slowly. However, the intervention's impact over a longer time horizon must be considered in order to make informed decisions about whether the change should be adopted. One long-run effect of treatment of particular interest is the effect of changes in user preferences on the outcome. If the intervention has some significant positive or negative impact on user experience that takes time to realize, there may be consequences that are not well-captured by immediate changes to the platform metrics. For example, due to a novelty effect, users may initially respond positively to a series of flashy UI changes by spending more time on the platform. But over time, as the users interact repeatedly with these changes, the changes may begin to irritate the users, and users may reduce usage or even abandon the platform.\footnote{Although ``novelty effects'' are sometimes used to refer to changes in user behavior, we prefer the term user learning since it captures user effects that have a long-term impact on outcomes, not just those which disappear after users have adjusted to the new setting.} 

\citet{hohnhold2015focus} introduced the dynamic Cookie-Cookie-Day (CCD) Experiment as a way to estimate and extrapolate user learning effects and long-term effects over time. For user learning, the CCD methodology relies on comparing the outcomes of individuals who are treated for the long-term, compared to individuals who are only treated for a single day, to estimate how the user learning effect accrues over time. However, this design requires that user learning is the only factor that can result in differences in the outcomes of long-term compared to short-term treated individuals in a system. In practice, there are other factors that affect long-term treated users differently. An important example of such a factor is personalization, which is present in a variety of online platforms that run randomized experiments to evaluate interventions. 

 Many online platforms are personalized based on a user's history of interactions with the platform. On a streaming service, previous views determine recommended movies. On an e-commerce website, previous purchases influence recommended products. This means that as an intervention changes  the users click patterns, for example, then the platform recommendations are affected, which feeds back into user clicks in the future. We show that if an online platform is personalized, then the CCD experiment estimates a combination of user learning and personalization effects, which are difficult to disentangle since they generally occur at different rates.

The primary goal of this paper is to propose new experimental designs that can isolate the user learning effect and personalization effect. There are two main reasons to isolate the user learning effect as opposed to directly estimating the total effect. Firstly, an empirical observation by \citet{hohnhold2015focus} is that we can be confident in extrapolating the user learning effect alone, but not necessarily the combined learning and personalization effects, and it is therefore better to have separate models for each. Secondly, the user learning estimand can in itself be important for evaluating interventions because the learning estimand is a good proxy measure of quality \cite{sadeghi2022novelty}.

We first develop a causal model of user behavior in an online platform, for binary treatments that affect the system that the user interacts with. Using a potential outcomes framework \citep{imbens2015causal}, the model formalizes how treatments affect individual behavior both directly through their immediate actions and indirectly through changes in their preferences over time and changes in the system's recommendations. This model leads to formal definitions of the user learning effect, the personalization effect, and the direct effect, the sum of which is the total causal effect of a treatment that is implemented for the long-term. 

Without additional intervention, the personalization of the system and changes in user preferences move together over time. In order to estimate the causal effect that is mediated through personalization and through user learning separately, we require designs that intervene in some way in the personalization system. We consider three types of designs. The first, which we call CCD-Switch, assigns a random cohort of users to be treated for the long-term but receive the same recommendations as similar users who remain in control. This creates a cohort of users whose preferences change in response to the treatment but whose personalization evolves as if they were in control. An alternative, simpler design based on this idea is CCD-Freeze, where a random cohort of users is treated but their personalization is based on their actions during the pre-experiment period, when they were not exposed to treatment. Under CCD-Switch and CCD-Freeze, both the user learning effect and personalization effect can be estimated with low bias. Lastly, we show that under stronger assumptions, when the personalization of a system is applied at the level of \textit{clusters} of similar users, then the standard CCD experiment can be used to estimate user learning effects.  We characterize the bias and discuss the relative benefits and drawbacks of each of these designs.

The paper is structured as follows. The remainder of Section \ref{sec:intro} reviews the related literature on experimental design for estimating treatment effects in the setting of online platforms where there may be interference between users and the effects may be dynamic. Section \ref{sec:potentialoutcomes} defines the direct, user learning, and personalization effects using the potential outcomes framework, and discusses why a standard CCD experiment cannot recover the user learning effect in a personalized system. In Section \ref{sec:design}, we define the three designs that we propose for measuring user learning in the presence of a personalized system. Section \ref{sec:estimation} discusses how medium-term estimates of the direct, user learning, and personalization effect can be extrapolated to the long-term by making parametric assumptions. In Section \ref{sec:empirical}, we include real-world evidence from a variety of CCD-type experiments in a large scale system that shows that the personalization effect is of meaningful magnitude and distorts the measurement of user learning effects. In Section \ref{sec:simulation}, we present empirical results comparing the different designs discussed through a simulation study which considers an intervention for a movie recommendation system. Additionally, for the setting where user learning converges very slowly over a period of months, we discuss how the user learning effect can be extrapolated from a relatively short experiment using a parametric assumption based on the empirical findings of \citet{hohnhold2015focus}. Concluding remarks can be found in Section \ref{sec:discussion}.

\subsection{Literature Review}
This paper is related to a variety of work addressing challenges in estimating treatment effects in online platforms using randomized experiments; see \citet{larsen2022statistical} and \citet{kohavi2020trustworthy} for a review and \citet{knijnenburg2015evaluating} for a discussion specific to recommendation systems. There is a growing literature that designs more complex randomized experiments. One strand of the literature addresses issues caused by interference, when individuals interact  with one another through an equilibrium, a network, or a market platform~\cite{fradkin2021reciprocity, eckles2016design}. There is a large literature on cluster-randomized designs, see \citet{hudgens2008toward} for a general cluster design, \citet{harshaw2021design} for design in two-sided markets, and \citet{leung2022rate} for designs under a spatial model. \citet{bajari2021multiple} and \citet{johari2022experimental} introduce  designs that randomize at an item-user level for two-sided markets, \citet{munro2021treatment} analyzes an individual-level augmented randomized experiment that jointly randomizes prices and treatments, and \citet{viviano2020experimental} studies two-wave experiments under network interference to estimate spillover effects. This literature usually studies estimands defined at a single point in time.

 Models where treatment assignment and outcomes can vary over time can also lead to more complex estimands and experimental designs. \citet{bojinov2021panel} discuss dynamic experiments in panels and introduce related estimands. For the statistical properties of switchback experiments, see \citet{bojinov2022design} and \citet{hu2022switchback} and for the design of time series experiments see \citet{bojinov2019time}. This paper is concerned with specific dynamic effect, the effect of a treatment over time that is mediated through user learning and personalization of an online system. This focus on experimental design for a dynamic mediated effect is new.  

There is a large literature on mediation analysis in observational and experimental data; an incomplete review includes \citet{pearl2022direct, vanderweele2016mediation} and \citet{imai2010identification}. This paper tackles a dynamic setting, where exposures and mediators vary over time, and the estimand is also defined over time, see \citet{vanderweele2017mediation}. This paper's contribution is in introduction of new experimental designs in a challenging dynamic setting, building on previous work on experimental design for mediation analysis in the static setting of \citet{imai2013experimental}. We are in the setting of two interacting mediators, and since each of the methods proposed in this paper involves an intervention with the goal of making personalization comparable in the control and treatment groups, our estimands can be viewed as a variation of controlled mediated effects, see \citet{vanderweele2011controlled}. We again acknowledge the work of~\citet{hohnhold2015focus}, who studied estimation of user learning from an experimental design perspective under the assumption of no-personalization. Related work from  \citet{sadeghi2022novelty} introduces an observational approach to measure user learning effects, also without personalization. \citet{miller2019targeted} discuss the impact on bias of targeted marketing, which motivates measuring personalization effects separately from a fairness perspective.

\section{Defining User Learning in a Potential Outcomes Model}  \label{sec:potentialoutcomes}

We consider a potential outcomes model where $ i \in \{ 1, \ldots n \}$ users interact with an online platform over time $t \in \{1, \ldots T\}$. There is a binary intervention $W_{it} \in \{0, 1\}$ that can be varied over time for each user. A user's actions at time $t$ are $X_{it} \in \mathcal X$ and their historical actions are defined as $\bm X_i^{<t}$. A user's actions may be high-dimensional, and include a variety of information about how long they interacted with the online platform and what links were clicked. The outcome of interest is a summary statistic of the user's actions, $Y_{it} = f(X_{it})$. For example, an outcome could be whether or not the user clicks on a recommended item when engaging with a recommendation system. The user has potential outcomes $Y_{it}(W_{it}, \bm W_{i}^{<t})$, so each possible history of user treatments and the current treatment they are exposed to can lead to a different user action at time $t$.  

We next describe a model of how these potential outcomes are generated as the system state (including personalization of the system) and user preferences are affected by a treatment over time. We assume that user actions are generated by the following process: at time $t$, a user with unobserved preferences $U_{it}$ views the system state $S_{it} \in \mathcal S$ and chooses a set of actions $X_{it} \in \mathcal X$.  We can represent a user's choice process as a function $g: \mathcal S \times \mathcal U \rightarrow \mathcal X$, so $X_{it} = g(S_{it}, U_{it})$.\footnote{In many choice models, conditional on a user's ``type'', there is still some randomness in the option chosen. We are choosing to define unobserved preferences to capture all the randomness in a choice process, so that conditional on $U_{it}$, actions taken are deterministic. This simplifies the exposition without affecting its generality} $S_{it}$ and $U_{it}$, and therefore $X_{it}$ as well, depend on the history of treatments; we describe this in more detail in the following paragraphs. 

The treatment is any intervention, such as a UI change or algorithm change, that directly impacts the system that the user interacts with.\footnote{ Some interventions, such as a change in which historical actions are used for personalization, may impact the parts of the system that are personalized. More often, interventions impact system features that are not personalized.} The system state might include what kind of layout a user sees, or what links they are recommended. The system state is personalized when it depends on the user's action history before time $t$. We capture the personalization of the system with a function $p(\cdot)$ that maps from the history of user actions to $\mathbb R^d$. For example, the recommendations that a user sees may depend on what kind of links they clicked in previous sessions. The system state $S_{it} \in \mathcal S$ for a given history of user treatments is then defined using a function $s: \{0, 1\} \times \mathbb R^d \rightarrow \mathcal S$, so $S_{it}(W_{it}, \bm W_i^{<t}) = s(W_{it}, p(\bm X_{i}^{<t} (\bm W_{i}^{<t}))$.  The first argument captures the immediate effect of the treatment on the system state, and the second argument captures the long-term effect of the treatment on the system state through historical user actions that affect personalization. The system state is not personalized when the system features that a user receives do not depend on their previous actions, and $p(x)$ is constant for any possible history of user actions.\footnote{ For the purposes of this paper, we assume that the personalization algorithm in the system is not updated during the evaluation period of the experiments that we introduce. Since the algorithm is not retrained on data generated by the experiments, then $S_{it}$ does not depend on the historical actions of user $j$, where $j \neq i$.}

\begin{figure}
\centering
\begin{tikzpicture}
    \node[state] (u) at (0,0) {$U^{}_{it}$};

     \node[state] (s) [above=of u]{$S^{}_{it}$}; 
    \node[state] (h) [left =of u]{$\bm W^{<t}_i$}; 
    \node[state] (x) [right =of u] {$X^{}_{it}$}; 
        \node[state] (y) [right =of x] {$Y^{}_{it}$};
     \node[state] (w) [right =of s]{$W^{}_{it}$}; 
    \node[state] (xh) [above =of h]{ $\bm X^{<t}_i$}; 
    
    \path (h) edge (xh); 
    \path (xh) edge (s); 
    \path (u) edge (x);
    \path (w) edge (s); 
    \path (h) edge (u); 
    \path (s) edge (x); 
    \path (x) edge (y); 
\end{tikzpicture}
\caption{Outcomes depend on user preferences and the system state. \label{graph} } 
\end{figure}
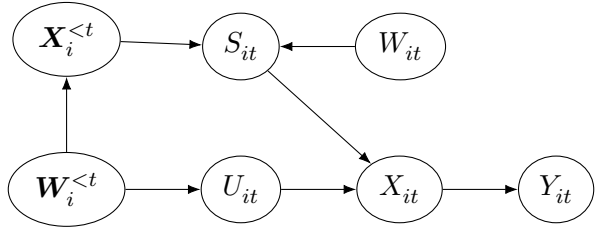

A user's preferences encompass a variety of unobserved factors that determine how a user makes choices when interacting with the platform; the preferences may evolve over time. Although we assume the intervention will not immediately impact user preferences, it may have a long-term impact on $U_{it}$ that realizes slowly as the user learns about their optimal set of actions in response to a platform change. We make explicit this dependence on historical treatments by writing $U_{it} = U_{it}(\bm W_{i}^{<t})$. In Figure \ref{graph}, we include a causal graph that summarizes this discussion of how outcomes depend on treatments through user preferences, the system state, and user actions. This discussion leads to the following notation for potential outcomes in our model: $Y_{it}(W_{it}, \bm W_{i}^{<t}) = Y_{it}(S_{it}(W_{it}, \bm W_{i}^{<t}), U_{it}(\bm W_{i}^{<t}))$. 

The platform has two complementary goals when designing an experiment. First, to evaluate how a binary treatment impacts outcomes over time when rolled out to all users. Second, to evaluate if the treatment is impacting user preferences over time in a way that affects outcomes---we call this a user learning effect.  We formalize these two estimands as $\tau^{TOTAL}_t$ and $\tau^U_t$ below using the potential outcomes that we have defined in the first part of this section. The platform is also often interested in evaluating long-term effects, where the time period of interest goes beyond the window in which an experiment is run. In Section \ref{sec:estimation}, we describe how a parametric assumption on the user learning effect, defined below, can be used for extrapolation.

The causal total impact of a treatment at a single point in time is the difference in outcomes for users that have been treated from time $1$ to $t$ compared to users who have not interacted with a treated system: 
\[ \tau^{TOTAL}_t
 = \frac{1}{n} \sum \limits_{i=1}^n \left [  Y_{it}(S_{it}(1, \bm 1^{<t}), U_{it}(\bm 1^{<t})) - Y_{it}(S_{it}(0, \bm 0^{<t}), U_{it}( \bm 0^{<t}))\right], \]
where $\bm 1^{<t}$ is a $t-1$ length vector of ones and $\bm 0^{<t}$ is a $t-1$ length vector of zeros. Without further assumptions, we can decompose the total impact of the treatment into three types of distinct causal effects: $\tau^{TOTAL}_t =  \tau^U_t + \tau^P_t + \tau^S_t$.

\begin{enumerate} 
\item \textbf{The personalization effect:} The effect on outcomes of long-term changes in the personalization system for a treated user, holding user preferences fixed. 
\[  \tau^P_t = \frac{1}{n} \sum \limits_{i=1}^n  \left [ Y_{it}(S_{it}(1, \bm 1^{<t}), U_{it}( \bm 1^{<t})) - Y_{it}(S_{it}(1, \bm 0^{<t}), U_{it}( \bm 1^{<t})) \right].\]
\item \textbf{The user learning effect:} The effect on outcomes of long-term changes in user preferences for a treated user, holding personalization fixed.
\[ \tau^U_t = \frac{1}{n} \sum \limits_{i=1}^n \left [ Y_{it}(S_{it}(1, \bm 0^{<t}), U_{it}( \bm 1^{<t})) - Y_{it}(S_{it}(1, \bm 0^{<t}), U_{it}( \bm 0^{<t}))\right].  \] 
\item \textbf{The direct effect:} The short-term impact on outcomes due to a change in the system state for users who have otherwise not been treated.
\[ \tau^S_t = \frac{1}{n} \sum \limits_{i=1}^n \left [ Y_{it}(S_{it}(1, \bm 0^{<t}), U_{it}(\bm 0^{<t})) - Y_{it}(S_{it}(0, \bm 0^{<t}), U_{it}( \bm 0^{<t}))\right]. \]
\end{enumerate} 

Understanding how user preferences are affected in the long-term by a treatment, leading to a change in user actions, is of independent interest. There are multiple definitions of user learning effects possible, that differ in how the treatment and the personalization state is held fixed as user preferences change due to the treatment. Above we have chosen a natural definition of the user learning effect that is amenable to estimation using the experimental designs introduced in Section \ref{sec:design} below. 

We conclude this section by discussing how two existing experimental designs relate to the causal estimands defined. First, a long-term A/B test that maintains one random group of individuals in treatment and one random group of individuals in control will provide an unbiased estimate of $\tau^{TOTAL}_t$. However, the user learning effect is not identifiable from a long-term A/B test. 

\begin{definition} \textbf{Long-Term A/B Test} 
\label{def:ab}
\begin{enumerate} 
\item Randomly split users into a Control Cohort ($C$) and a Treated Cohort ($T$) 
\item For $t \in \{1, \ldots, T\}$ periods, treat individuals in the treated cohort and maintain the rest of the users in control. 
\end{enumerate} 
\end{definition}

Let $G_{it} \in \{C, T \} $ capture the cohort that an individual is randomly assigned to at time $t$.

\begin{proposition}\label{prop:ab}
Under Definition \ref{def:ab}, 
\[\hat \tau_t = \frac{1}{n} \sum \limits_{i=1}^n \frac{\mathbbm{1}(G_{it} = T)}{ Pr(G_{it} = T)} Y_{it} -\frac{1}{n} \sum \limits_{i=1}^n \frac{\mathbbm{1}(G_{it} = C)}{ Pr(G_{it} = C)} Y_{it} \]  is an unbiased estimate of the effect $\tau^{TOTAL}_t$. 
\end{proposition}  

Proposition \ref{prop:ab} indicates that in each time period $t=1, \ldots, T$ during the long-term A/B test, we can make a noisy measurement of $\tau^{TOTAL}_t$. There are two ways in which the measurement of $\tau^{TOTAL}_t$ from an A/B test is incomplete. First, we we may be interested in $\tau^{TOTAL}_s$ for $s >T$. In Section \ref{sec:estimation}, we discuss how estimating components of $\tau^{TOTAL}_t$ separately can be useful for extrapolation. Second, we cannot evaluate the user learning effect separately using data from the A/B test only.

\citet{hohnhold2015focus} introduces the Cookie-Cookie-Day (CCD) experiment design for estimating a time series of user learning effects, which we describe in Definition \ref{def:ccd} below. Here ``Cookie'' refers to CT, i.e.  an experiment arm consisting of a cohort of browser cookies (proxies for users) which receive treatment throughout the experiment. ``Cookie-Day'' refers to an experiment arm whose underlying browser cookies change every day, specifically each day a random sample of cookies are selected to receive treatment for that single day from a large pool of cookies (the rest of which receive control). We denote the Cookie-Day arm by $CDT_t$ below. CCD experiments compare the Cookie arm to the Cookie-Day arm. The Cookie-Day arm plays the role of the control arm when measuring user learning effects because the underlying users should have little or no previous exposure to the treatment, and therefore have not accumulated user learning, whereas the users in the Cookie arm have. As before, the difference in the $CT$ and $CC$ cohort is an unbiased estimate of $\tau^{TOTAL}_t$. Meanwhile, the $CDT_t$ cohort allows estimation of $\tau^S_t$ and also $\tau^U_t$ in the absence of personalization. The CCD design is illustrated in the first half of Figure \ref{fig:designs}.

\begin{definition} \label{def:ccd}\textbf{CCD Experiment.} 
\begin{enumerate} 
\item Randomly split users into a Cookie Cohort ($C$) and a Cookie-Day Cohort ($CD$) 
\item Treat a random fraction of the cookie cohort ($CT$) for $T$ periods and maintain the rest in control ($CC$) 
\item Within the Cookie-Day Cohort, assign individuals randomly to be treated for one day only for $t \in \{1, \ldots T\}$ ($CDT_t$). Those who are in the $CD$ cohort at time $t$ but are not treated are labelled $CDC_t$. 
\end{enumerate} 
\end{definition} 

We use the variable $G_{it} \in \{CT, CDT_t, CC \}$  to capture the cohort that an individual is randomly assigned to. We do not use the $CDC_t$ cohort to measure user learning. When the system state is not personalized, we can use the CCD experiment to estimate the user learning effect by comparing the average outcomes of individuals in the $CT$ cohort compared to the $CDT_t$ cohort~\cite{hohnhold2015focus}:
 
 \[ \hat \tau^{CCD, U}_t = \frac{1}{n} \sum \limits_{i=1}^n \frac{\mathbbm{1}(G_{it} = CT)}{Pr(G_{it} = CT)} Y_{it} - \frac{1}{n} \sum \limits_{i=1}^n \frac{\mathbbm{1}(G_{it} = CDT_t)}{Pr(G_{it} = CDT_t)  } Y_{it} .  \]  

\begin{proposition} \label{imposs}
Under Definition \ref{def:ccd}, $\hat \tau^{CCD, U}_t$ is an unbiased estimate of $\tau^U_t + \tau^P_t$. The differences in mean outcomes between the $CDT_t$ and the $CC$ cohort, 
\[ \hat \tau^{CCD, S} = \frac{1}{n} \frac{1}{Pr(G_{it} = CDT_t)} \sum \limits_{i=1}^n \mathbbm{1}(G_{it} = CDT_t) Y_{it} - \frac{1}{n} \frac{1}{ Pr(G_{it} = CC) }  \sum \limits_{i=1}^n \mathbbm{1}(G_{it} = CC) Y_{it}, \]
is an unbiased estimate of the direct effect $\tau^S_t$. 
\end{proposition} 

When there is no personalization, such that the system state does not depend on past user actions, then $\tau^P_t=0$, and the CCD experiment measures the user learning effect $\tau^U_t$. When the system is personalized based on the history of user actions, then the difference between the $CT$ and $CDT_t$ cohorts measures the sum of the user learning effect and personalization effect. This means that evaluating user learning effects based on a CCD experiment in personalized systems can be misleading. As described in more detail using an example in the next section, it is possible that a CCD experiment measures a positive effect, but the true user learning effect is zero, or even negative. In Section \ref{sec:design}, we suggest new experimental designs that can measure user learning effects in the presence of personalization.

\subsection{Example of incorrect user learning measurements due to personalization}\label{sec:example}
The Cookie-Cookie-Day experiment is designed to measure long-term user learning by comparing two cohorts of users who receive the treatment, but who differ in their treatment history. The CCD method assumes that, once we control for the direct effect of the treatment by assigning treatment to both CT and $CDT_t$ users, any remaining difference in users' responses is the result of user learning. Unfortunately, this assumption is not always true. Many recommendation systems are \textit{personalized}, so that the items shown to users differ based on their past actions, which in turn depends on the user's treatment history (see Figure \ref{graph}). 

Personalization causes systematic differences between the recommendations made to CT and $CDT_t$ cohorts. In a CCD experiment, these systematic differences are conflated with differences in behavior due to user learning, resulting in incorrect measurements of user learning. The following example illustrates this problem.


\example{\textbf{Annotating award-winning movies} \label{ex:movies}

Suppose a streaming movie service seeks to increase user engagement by highlighting award-winning movies with a special annotation.
The service believes the annotations will have the \textit{direct effect} of increasing user \textit{watch rate}, measured as the fraction of recommended movies that the user chooses to watch.
The service also hypothesizes that there will be a \textit{long-term user learning effect} of increased watch rate over time, as users learn that the service recommends high-quality movies and engage more frequently with its recommendations.
To measure long-term user learning, the service runs a Cookie-Cookie-Day experiment, as described by \citet{hohnhold2015focus}. Figure \ref{fig:movie-rec-1} illustrates the example.

\begin{figure}
    \centering
    \includegraphics[clip, trim=0cm 9.2cm 4.5cm 0cm, width=\textwidth]{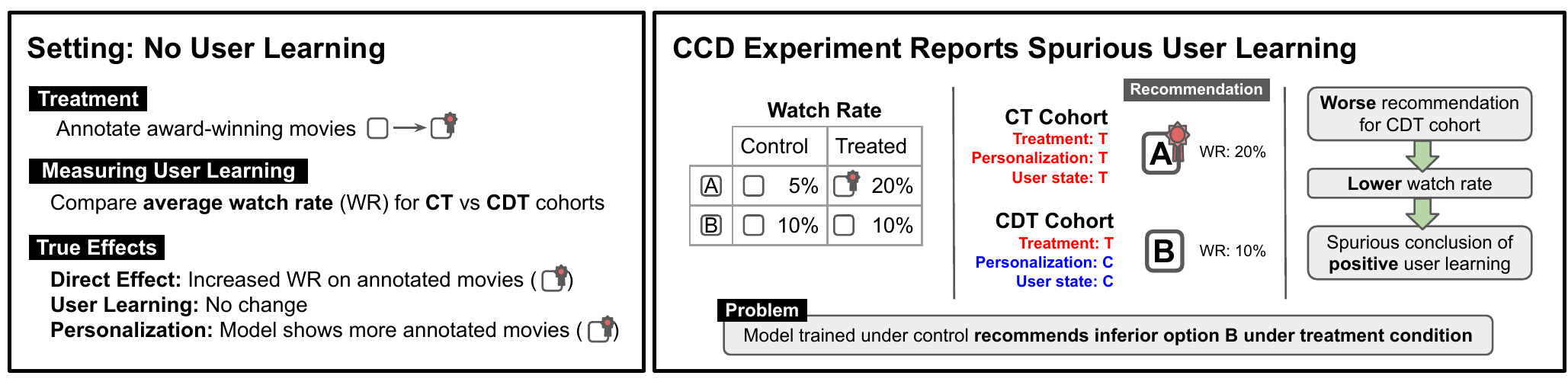}
    \caption{Illustration of the CCD experiment's biased user learning estimates in the presence of system personalization. In this example, movie B is not eligible for an award; it has a watch rate of 10\% under both treatment and control. Movie A is eligible for an award, with a watch rate of 5\% under control and 20\% under treatment. The personalized recommendation system learns the true watch rate for the CT cohort, and recommends movie A to CT users. However, the system learns CDT users' preferences under the control, and therefore incorrectly recommends movie B to CDT users. This difference in recommendation causes CDT users to watch recommended movies at a lower rate than CT users, leading to a spurious conclusion of positive user learning. 
    }
    \label{fig:movie-rec-1}
\end{figure}

The service uses personalized recommendations to suggest movies to user $i$ based on which movies $\textbf{X}_i^{<t}$ the user has watched in the past, which is in turn affected by the user's past treatments $\textbf{W}_i^{<t}$ (see Figure \ref{graph}).
Because of this causal link from the user's past treatment to the system's personalization,
users in the CD cohort will receive systematically different personalized recommendations from users in the CT cohort. 

Consider a setting in which there is no user learning; $\tau^U = 0$. Users in the CT and CDT cohorts have identical outcomes when provided the same suggestions $S$; i.e. $Y_{it}(S_{it},U_{it}) = Y_{it}(S_{it}, U_{it}')$.
Because suggestsions are personalized, users in the CT and CDT cohorts have systematically different histories, and therefore are provided systematically different recommendations.
In this example, the treatment increases the watch rate of award-winning movies, so CT users will watch these movies at a higher rate than CDT users. The system will learn to serve award-winning movies to CT users at a higher rate than CDT users, but since both CT and CDT users are treated, CDT users see recommendations that do not match their true preferences. This leads to decreased watch rate in the CDT cohort as compared with the CT cohort, leading to a spurious conclusion of positive user learning.

}

\subsection{Two types of errors}
We have seen that personalization can interfere with the results of CCD experiments, leading to spurious conclusions of positive user learning when in fact no user learning has occurred. We also observe that personalization can cause spurious conclusions of negative user learning; see Appendix \ref{app:example} for an example. In general, the effects of personalization combine with the effects of user learning in a CCD experiment to produce biased estimates of user learning. If the bias is toward positive user learning, then the platform may launch a feature that actually harms its users, despite believing the feature was beneficial. On the other hand, if the bias is toward negative learning, a feature that is truly beneficial may fail to launch.


Since system personalization biases the estimate of user learning in a CCD experiment, a natural solution is to remove personalization when measuring user learning. Instead of personalizing recommendations based on a user's watch history, the system could recommend the same set of movies to every user. Recommendations would be the same for $CT$ and $CDT_t$ users, removing the bias due to personalization. Unfortunately, removing personalization is undesirable in many settings, as it significantly degrades the quality of the user's experience. Instead, we seek experimental designs that enable us to measure user learning in the presence of personalization. The next section describes such designs.




\section{Experimental Design} \label{sec:design}

\begin{figure}
    \centering
    \includegraphics[clip, trim=0cm 7.8cm 0cm 0cm, width=\textwidth]{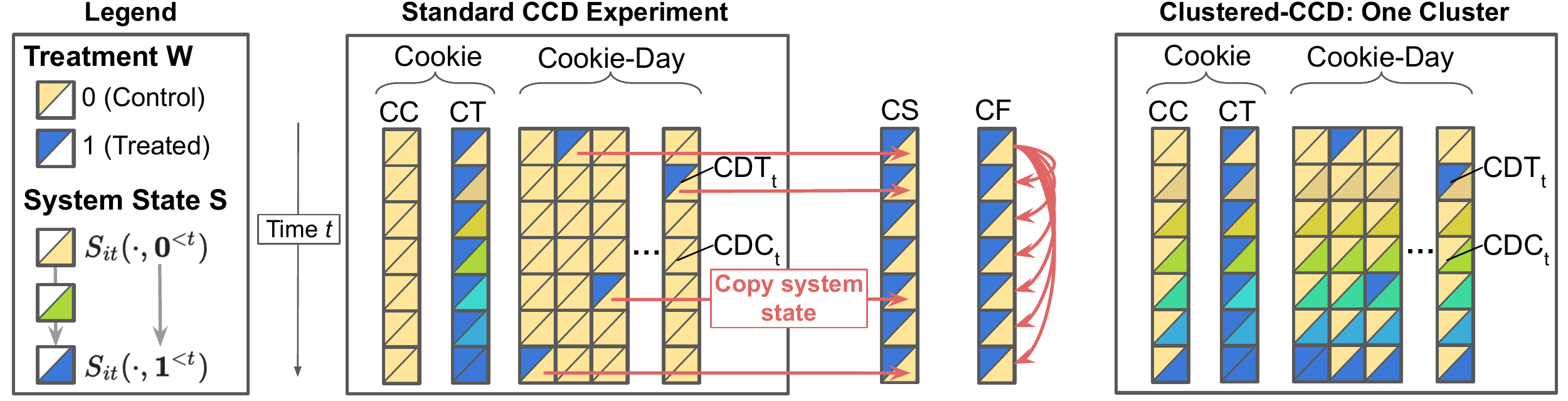}
    \caption{Illustration of the cohorts used in the standard CCD experiment and in our proposed designs. As the CT units are exposed to treatment over time, their system state evolves from ``yellow'' to ``green'' to ``blue''. CS corresponds to the Cohort-Switch mechanism, CF corresponds to the Cohort-Freeze mechanism.
    }
    \label{fig:designs}
\end{figure}

In the first part of this section, we introduce designs that manipulate the individual-level personalization of the system for certain individuals in the experiment. This leads to estimators of both the personalization effect and the user learning effect. In the second part of this section, we consider designs that cluster personalization, serving the same system state to a subgroup of users with similar user features. Under stronger assumptions, the clustering designs lead to estimators of the user learning effect, but not the personalization effect of a treatment. The proposed designs are illustrated in Figure \ref{fig:designs}.

The first design we introduce is the CCD-Switch Experiment, described in detail in Design \ref{def:ccds}. In order to disentangle $\tau^U_t$ from $\tau^{TOTAL}_t$ in a system without personalization, the CCD experiment adds an additional experimental cohort to the long-term A/B test. For systems with personalization, we can estimate $\tau^U_t$ by adding an additional cohort to the vanilla CCD experiment. Each individual in the switch cohort is treated for the long-term, such that user learning occurs as their preferences evolve. However, they interact with a treated system that is personalized based on a similar user's actions who has been in control for the duration of the experiment. This leads to a good proxy for the outcomes of a user that is learned in response to the treatment but still views a system that is personalized as if they were behaving under control conditions. 

\begin{definition}\label{def:ccds} \textbf{CCD-Switch Experiment.}
\begin{enumerate} 
\item Split users randomly into a Cookie Cohort ($C$), a Cookie-Day Cohort ($CD$), and a Switch-Cohort $(CS)$.
\item For $T$ periods, treat  a random fraction of the Cookie Cohort ($CT$), maintaining the rest of the Cookie Cohort in control ($CC$) . 
\item Within the Cookie-Day Cohort, assign individuals randomly to be treated for one day only for $t \in \{1, \ldots T\}$ ($CDT_t$). Those who are in the $CD$ cohort at time $t$ but are not treated are labelled $CDC_t$. 
\item  For $T$ periods, treat the Switch-Cohort ($CS$) but match each individual $i$ in $CS$ to the individual $j(i)$ in cohort $CDT_t$ with the closest pre-experiment user action history and serve $S_{it}(1, p(X_{j(i)}^{<t}))$ to user $i$ rather than $S_{it}(1, \bm 1^{<t})$. 
\end{enumerate} 
\end{definition} 

In the description of the CCD-Switch experiment, we describe using a control user's action history to determine the personalization of individual $i$. We can also match to multiple other users rather than a single user, and determine the personalization of individual $i$ based on the average personalization of the multiple matched users. 

Let $G_{it}$ be a discrete variable that represents the treatment cohort of individual $i$. As in the CCD experiment, we can estimate $\tau^S_t$ by comparing the $CDT_t$ and the $CC$ cohort and $\tau^{TOTAL}_t$ by comparing the $CT$ and $CC$ cohorts. We can now also disentangle $\tau^U_t$ and $\tau^P_t$. The difference in average outcomes between the $CS$ and the $CDT_t$ cohorts is a good estimate of user learning: 
 \[ \hat \tau^{DS, U}_t = \frac{1}{n}   \sum \limits_{i=1}^n \frac{ \mathbbm{1}(G_{it} = CS) }{Pr(G_{it} = CS)} Y_{it}-  \frac{1}{n} \sum \limits_{i=1}^n \frac{\mathbbm{1}(G_{it} = CDT_t)}{ Pr(G_{it} = CDT_t) } Y_{it}.  \]  
The difference in averages outcomes between the $CT$ and the $CS$ cohort is a good estimate of the personalization effect: 
 \[ \hat \tau^{DS, P}_t = \frac{1}{n}   \sum \limits_{i=1}^n \frac{ \mathbbm{1}(G_{it} = CT) }{Pr(G_{it} = CT)} Y_{it}-  \frac{1}{n} \sum \limits_{i=1}^n \frac{\mathbbm{1}(G_{it} = CS)}{ Pr(G_{it} = CS) } Y_{it}.   \]

\begin{proposition} \label{switch} Under Definition \ref{def:ccds}, $ \hat \tau^{DS, U}_t = \tau^U_t + b^{DS}_t$ and $\hat \tau^{DS, P}_t = \tau^P_t - b^{DS}_t$. The bias term is: 

\[ b^{DS}_t = \frac{1}{n} \left[ \sum \limits_{i=1}^n Y_{it}(s(1, p(\bm X_{j(i)}^{<t})), U_{it}(1, \bm 1^{<t})) - Y_{it}(S_{it}(1, \bm 0^{<t}), U_{it}(\bm 1^{<t}))\right], \]

where $s(1, p(\bm X_{j(i)}^{<t}))$ is the system state for the matched user in the $CDT_t$ cohort. 

\end{proposition}
The bias term is usually non-zero since we cannot guarantee that the actions of the matched user $j$ in $CDT_t$ cohort are exactly the same as user $i$'s unobserved choices if they were controlled, even if their pre-experiment actions are nearly identical. As the sample size grows larger, we would expect that matching would improve and the bias term would reduce. 

As the bias term in the CCD-Switch experiment illustrates, the ideal measure of long-term user learning measures the outcome of users that are treated but have the personalization as if they were in the control condition up to time $t$. CCD-Switch approximates this counterfactual personalization state using the personalized state of some treated user(s) whose history approximates the counterfactual history of a given treated user. A different proposal is to use the treated user's pre-experiment personalization state as a proxy for its counterfactual controlled personalization state, a design we call CCD-Freeze.
In this design, described below, there is an additional Frozen-Cohort ($CF$), who are treated but receive system personalization based on their fixed pre-experiment user action history, which we call $\bm X_{i}^*$. The resulting system state received by the freeze cohort is $S_{it} = s(W_{it}, p(\bm X_{i}^*))$.

\begin{definition} \label{def:freeze}\textbf{CCD-Freeze Experiment.}
\begin{enumerate} 
\item Split users randomly into a Cookie Cohort ($C$), a Cookie-Day Cohort ($CD$) and Frozen-Cohort $(CF)$. For the Frozen-Cohort, fix (``freeze'') their user features used for personalization at the pre-experiment values.
\item Treat the Freeze Cohort (CF) for $T$ periods. 
\item For $T$ periods, treat a random fraction of the Cookie Cohort ($CT$) and maintain the rest in control ($CC$). 
\item Within the Cookie-Day Cohort, assign individuals randomly to be treated for one day only for $t \in \{1, \ldots T\}$ ($CDT_t$). Those who are in the $CD$ cohort at time $t$ but are not treated are labelled $CDC_t$. 
\end{enumerate} 
\end{definition} 

If the pre-experiment user action history is a good proxy for their actions under control, then we would expect $s(1, p(\bm X_{i}^*))$ to be close to $S_{it}(1, \bm 0^{<t})$ and the following to be a good estimate of $\tau^U_t$: 

\[ \hat \tau^{CCF, U}_t = \frac{1}{n}   \sum \limits_{i=1}^n \frac{ \mathbbm{1}(G_{it} = CF) }{Pr(G_{it} = CF)} Y_{it}-  \frac{1}{n} \sum \limits_{i=1}^n \frac{\mathbbm{1}(G_{it} = CDT_t)}{ Pr(G_{it} = CDT_t) } Y_{it}.  \]

We can also estimate $\tau^P_t$: 

\[ \hat \tau^{CCF, P}_t =  \frac{1}{n}   \sum \limits_{i=1}^n \frac{ \mathbbm{1}(G_{it} = CT) }{Pr(G_{it} = CT)} Y_{it}-  \frac{1}{n} \sum \limits_{i=1}^n \frac{\mathbbm{1}(G_{it} = CF)}{ Pr(G_{it} = CF) } Y_{it}. \]

Deciding between running the CCD-Switch compared to the CCD-Frozen designs is context specific. In settings with a large sample where we would expect a lot of drift in user actions over time under a constant treatment status, then the CCD-Switch design may be preferable. In settings where it is operationally challenging to deliver the personalization of individual $j$ to individual $i$ and where we expect that the personalization delivered to a user in the pre-experiment period is a good proxy for the during-experiment period, then the CCD-Frozen design is preferable.

In the next section, we explore how we can return to a standard CCD design to estimate the user learning effect in environments where the personalization of the system is clustered, so that sub-groups of individuals receive the same recommendation, for example. 

\subsection{Clustered-CCDs} \label{sec:clustering}
In this section, we focus on adaptations to the existing CCD design that enable us to measure user learning without introducing additional cohorts. Recall that the central problem in CCD experiments with personalization is that the user's treatment history affects their personalized recommendations, which in turn affect the user's outcomes. This causal chain is illustrated in Figure \ref{graph}, and its consequences for CCD estimation of user learning are described in the example of Section \ref{sec:example}.

We propose to reduce the influence of a user's treatment history on their own personalization state by instead personalizing a user's recommendations according to the historical behavior of a \textit{cluster} of users. We call this the \textit{clustered-CCD design}.

\begin{definition} \label{def:lookalikes} \textbf{Clustered-CCD Experiment.}
\begin{enumerate} 
\item Assign each user to one of $K$ clusters of equal size, fixed throughout the experiment. Modify the recommendation algorithm so that recommendations are personalized based on the history of the \textit{cluster}, not the individual user. $S_{it} = s(W_{it}, p(\bm X_{c(i)}^{<t}))$, where the personalization depends on the action history and treatments of all individuals in the cluster $c(i)$. The rest proceeds like a standard CCD experiment.
\item Randomly split users into a Cookie Cohort ($C$) and a Cookie-Day Cohort ($CD$) 
\item Treat a random fraction of the cookie cohort ($CT$) for $T$ periods and maintain the rest in control ($CC$) 
\item Within the Cookie-Day Cohort, assign individuals randomly to be treated for one day only for $t \in \{1, \ldots T\}$ ($CDT_t$). Those who are in the $CD$ cohort at time $t$ but are not treated are labelled $CDC_t$. 
\end{enumerate} 
\end{definition} 

Then, our measure of user learning at time $t$, $\hat \tau^{L, U}_t$, as in the standard CCD experiment, is the difference in outcomes $Y_{it}$ between individuals in the CT cohort and the $CDT_t$ cohort at time $t$: 

\[ \hat \tau^U_L = \frac{1}{n}\sum \limits_{i=1}^n \frac{\mathbbm{1}(G_{it} = CT)}{Pr(G_{it} = CT)} Y_{it} - \frac{1}{n} \sum \limits_{i=1}^n \frac{\mathbbm{1}(G_{it} = CDT_t)}{Pr(G_{it} = CDT_t)  } Y_{it}\]

The Clustered-CCD approach of Definition \ref{def:lookalikes}  allows us to retain some personalization while keeping the distribution of personalization the same for the control and treatment users within the same cluster of any experiment (the treatment is assigned independently of the clusters). We are able to estimate the user learning effect under an additive separability assumption (defined below), and the direct effect by comparing the $CDT_t$ cohort to the $CC$ or $CDC_t$ cohort. Since the personalization state is the same for treated and control users, we are not able to estimate the personalization effect or the total effect from a clustered-CCD experiment. Running a separate long-term unclustered A/B test could be used to estimate the total effect and direct effect, and combined with the user learning estimates from the clustered experiment, the personalization effect can then be estimated. 

One benefit of the clustered CCD experiment is there is a privacy advantage of assigning personalization at the cluster level because it means that an individual user will be harder to identify from their personalization state. Additionally, in some cases, clustering provides a utility gain because some of the user features from which personalization states are computed may be very sparse, and borrowing information from similar users allows this data sparsity problem to be reduced. This will become apparent in Section~\ref{sec:simulation}. Furthermore, compared with the CCD-Switch  discussed in Definition \ref{def:ccds}, the Clustered-CCD method may be preferred in practice because it does not require the extra cohort needed for CCD-Switch---thereby improving experimental power---or the implementation of switching personalization states of the system. Additionally, in many practical settings, Clustered-CCD will also be preferred to the CCD-Freeze method of Definition \ref{def:freeze} because the key assumption of CCD-Freeze may not be met: personalization typically changes over time and so the pre-experiment state is not always a good proxy of the during-experiment state. 
Finally, both the CCD-Freeze and CCD-Switch methods require allocating a cohort whose recommendations are made according to a personalization state that does not match their treatment status; this degrades the quality of those users' recommendations, as discussed in Section~\ref{sec:simulation}. 


 The only restriction we place on the clusters is that (1) the treatment assignment must be independent of the cluster assignment (2) cluster assignments cannot be updated over the course of the experiment based on user actions during the experiment. Such dependent assignments are not allowed because they may cause a user's personalization to depend on its treatment assignment. For example, if the treatment caused users to take a particular action, they might be preferentially clustered together, leading to clusters that were mostly treated or mostly controlled. The personalization status of a user would once again be affected by its treatment status, with treated users more likely to see recommendations that lead to the treatment-specific action, as in the example of Section \ref{sec:example}.

While any fixed clustering that is assigned independently of the treatments will work for clustered-CCD, the ideal clustering identifies users with similar histories or personalization states. Clustering similar users means that recommendations provided on the basis of the cluster history are highly relevant to the individuals in the cluster, instead of being generic recommendations that are of little relevance to the user. We discuss how clusters might be generated in Appendix \ref{app:cluster}. 


We show that clustered-CCD experiments can be used to estimate the long-term user learning effect. With clustering, an individual's personalized state  now depends on their cluster assignment $c(i)$ and the action history of all individuals in their cluster. We represent this by the notation $S_{it} = s(W_{it}, P_{c(i)}(\bm W_{i}^{<t}))$ for the clustered setting, where $P_c$ depends on the action history (and treatments) for other users $j \neq i$ that are part of individual $i$'s cluster, as well as individual $i$'s action history.

\begin{assumption} \textbf{Additive Separability} \label{sep} 

For any two system states $s, s' \in \mathcal S$, 
\[  Y_{it}(s, U_{it}(\bm 1^{<t})) - Y_{it}(s, U_{it}( \bm 0^{<t})) = Y_{it}(s', U_{it}( \bm 1^{<t})) - Y_{it}(s', U_{it}( \bm 0^{<t})). \]
\end{assumption} 

\begin{proposition} \label{cluster} Under the clustered CCD experiment of Design \ref{def:lookalikes} and Assumption \ref{sep}, then 
\[ \hat \tau^{L, U}_t = \tau^U_t + b^L_t \]
where \[b^L_t =\frac{1}{n} \sum \limits_{i=1}^n \mathbb E[Y_{it}(s(1, \bm 1^{<t}, \bm W_{-i, c}^{<t}), U_{it}(\bm 0^{<t})) - Y_{it}(s(1, \bm 0^{<t}, \bm W_{-i, c}^{<t}), U_{it}(\bm 0^{<t}))] \] depends on how much a single individual's historical actions affect their cluster's personalization state. The expectation is taken over the cohort assignment of other individuals in each cluster. 
\end{proposition} 


One downside of the clustered-CCD design is that it relies on Assumption \ref{sep}, which states that the user learning and personalization effects are additively separable. For certain environments where there are interactions between learning and personalization, this assumption may not hold. Clustering also introduces interference between individuals, although under Assumption \ref{sep} the interference does not introduce additional bias in the estimate of the user learning effect; see Appendix~\ref{sec:interference} for a brief discussion.

Under Assumption \ref{sep}, the bias term $b_t^L$ in Proposition \ref{cluster} is the result of user $i$'s treatment affecting the recommendations $i$ sees through their history of actions. In the standard CCD experiment, which can be interpreted as a clustered-CCD design where each user belongs to its own cluster, a user's treatment history directly affects their personalization, and $b^L_t = \tau^P_t$. We suggest two aspects of clustered-CCD designs that can mitigate this issue. 

\paragraph{Larger Clusters.} As the number of users per cluster gets larger, the influence of any one user's history on the personalization state of its cluster diminishes. This in turn reduces the correlation between the user's treatment status and their personalization, reducing the bias in the clustered-CCD experiment.

\paragraph{Leave-one-out Serving.} In the leave-one-out serving approach, user $i$ is served recommendations that are personalized using the histories of all other users in its cluster except itself, $j\in c(i),~j\neq i$. This prevents user $i$'s history from directly impacting its recommendations, reducing the bias. Leave-one-out serving is not a complete solution because user $i$'s treatment still affects user $j$'s personalization, which in turn affects $j$'s actions and feeds back to user $i$'s personalization. However, leave-one-out serving is expected reduce the correlation between $i$'s treatment history and personalization.

\section{Extrapolation to Long-Term Causal Effects} 
\label{sec:estimation} 

In Section \ref{sec:design}, we introduced a variety of different experimental designs that extend the basic CCD experiment and provide noisy measurements of  the user learning effect $\tau^U_t$, the personalization effect $\tau^P_t$, and the direct effect $\tau^S_t$. The evolution of these time series during the experimental period is of interest in itself. However, it is also useful to consider how the time series estimated during the experiment can be used to extrapolate causal effects for time periods longer than experiment duration. This is of interest because user learning takes a very long time to converge. In the advertising setting, \citet{hohnhold2015focus} found that changes in user preferences induced by an intervention took many months to converge to a new steady state. They found that the user learning effects across a variety of different experiments follows an exponential curve, with $\tau^U_t = \alpha ( 1 - e^{-\beta t}) + e^U_t$.  $\beta$ was similar across different interventions while $\alpha$ varied. Estimating a parametric model for $\tau^U_t$ based on limited observations during an experiment can be used to extrapolate the expected learning effect as $t$ grows large. We can extend this approach to the setting with personalization by making parametric assumptions separately on $\tau^U_t$, $\tau^P_t$ and $\tau^S_t$. 

\begin{assumption}\label{as:par} \textbf{Parametric Structure.} 
The user learning effect, the personalization effect, and the direct effect have a parametric structure.
\begin{equation*}
\begin{split}
 \tau^U_t &= f^U(t, \beta^U) + \epsilon^U_{tn},\\
 \tau^P_t &= f^P(t, \beta^P) + \epsilon^P_{tn},\\
 \tau^S_t &= f^S(t, \beta^S) + \epsilon^S_{tn}, 
\end{split}
\end{equation*} 
where $\epsilon^q_{tn}$ are mean-zero random noise, $\mathbb E[\epsilon^q_{tn}] = 0$ for $q \in \{U, P, S\}$. 
\end{assumption}

In most settings, the direct effect would be expected to be constant over time, so that $f^S(t, \beta^S) = \beta^S$. In the simulation in Section \ref{sec:simulation}, we follow \citet{hohnhold2015focus} by choosing an exponential learning form for the user learning effect for $f^U_t(\cdot)$. We make a similar exponential learning assumption for $f^P_t(\cdot)$. In general the user learning and personalization effects accrue at different rates, so it is preferable to make a separate parametric assumptions for $\tau^U_t$ and $\tau^P_t$. 

Using $\hat \tau^U_t$, $\hat \tau^S_t$ and $\hat \tau^P_t$ we can estimate $\beta^q$ for $q \in \{U, P, S\}$ using a non-linear least squares procedure. 
\[ \hat \beta^q = \arg \min_{\beta^q} \frac{1}{T} \sum \limits_{t=1}^T (\hat \tau^q_t - f^q(t, \beta^q))^2. \]
Using the estimated parameters $\hat \beta^q$ we can estimate $\tau^q_s$ for time periods after the experiment, with $s >T$, as 
\[ \tilde \tau^q_s =  f^q(s, \hat \beta^q).\]

Another way that parametric assumptions can be useful for extrapolation is through a surrogacy approach \citep{athey2019surrogate}. The exponential learning form used for the user learning effect in \citet{hohnhold2015focus} converges to $\alpha$ as $t$ grows to infinity. Lets assume that the parametric form for each of the effects converges to a fixed value as $t \rightarrow \infty$, then under Assumption \ref{as:par}, we can write the limit of the expectation of the total causal effect of the treatment as: 
\begin{equation*}  
\begin{split}
\lim \limits_{t \rightarrow \infty} \mathbb E[ \tau^{TOTAL}_t ] &= \lim \limits_{t \rightarrow \infty} f^U(t, \beta^U) + f^S(t, \beta^S) + f^P(t, \beta^P) \\ 
& = \tau^U_* + \tau^P_* + \tau^S_*.
\end{split}
\end{equation*} 
\citet{hohnhold2015focus} provide evidence that certain short-term metrics evaluated using an A/B test are correlated with the long-term user learning effect, $\tau^U_* = \alpha$. This implies that a platform designer can run the modified CCD experiments of Section \ref{sec:estimation} for a variety of different interventions, estimate the parameters of $\beta^U, \beta^S, \beta^P$ for each intervention, and then investigate whether there are short-term metrics, estimable using a short-term A/B test, that are predictive of the intervention-specific parameter estimates. This surrogacy-type approach combined with the parametric model for each type of effect can speed up the development cycle for future interventions, by providing an indication of the estimated long-term user learning, personalization, and direct effect, without running a CCD experiment. 

\section{Empirical Evidence of Personalization Affecting User Learning Measurements}
\label{sec:empirical}

\begin{figure}
    \centering
    \includegraphics[scale=.5]{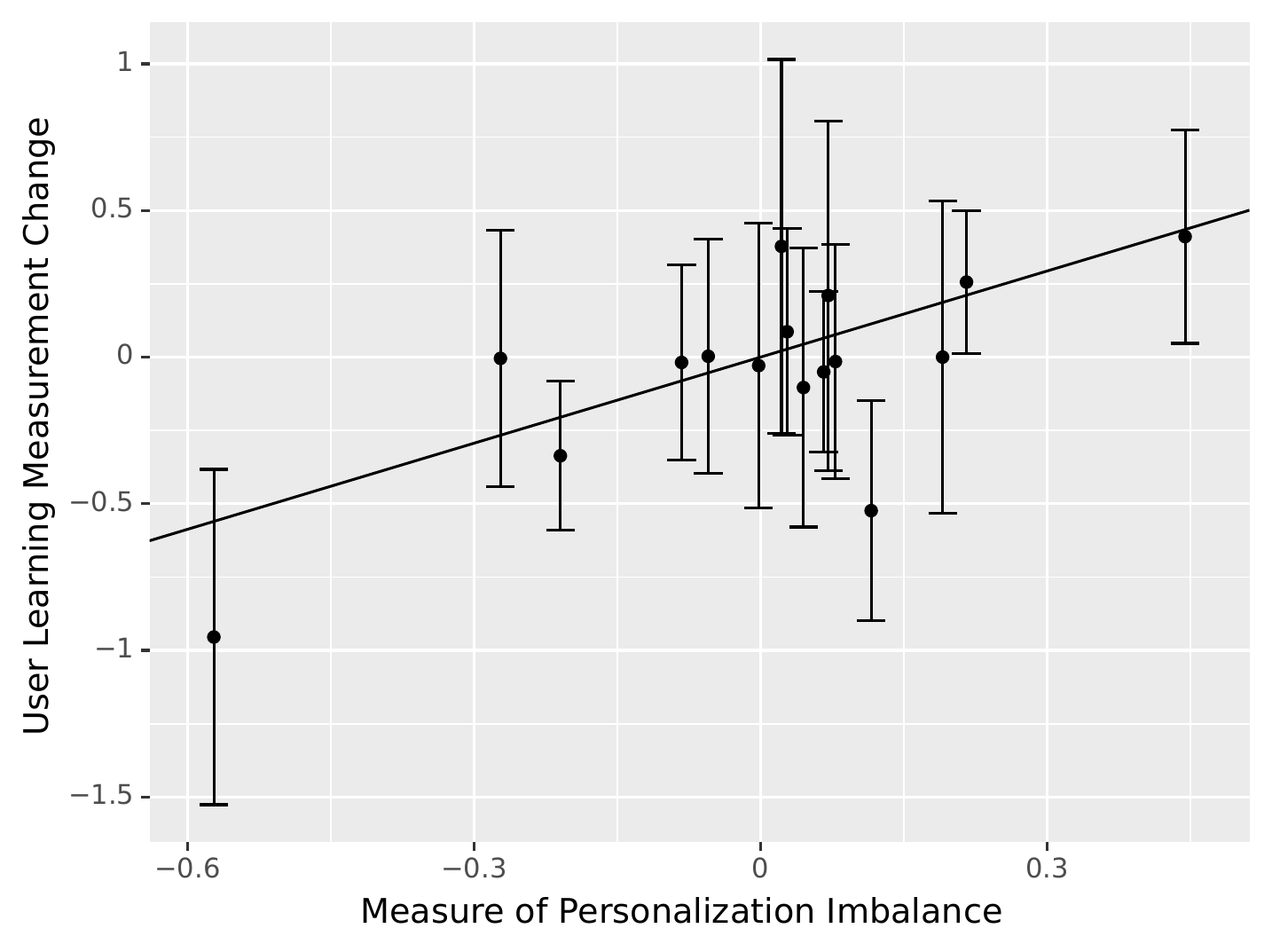}
    \caption{{Data collected from CCD interventions run on a Google Ads recommendation system}}
    \label{fig:example_real_world_data}
\end{figure}

We now demonstrate empirically the bias that personalization introduces to user learning measurements in standard CCD experiments---when the approaches proposed in Section \ref{sec:design} above are not used---as anticipated by~\citet{hohnhold2015focus}. We draw our results from a selection of CCD studies run by an  ads recommendation system at Google. We compare two sequential measurements of user learning: one \emph{with personalization} $\hat \tau^{U, WP}_{t_0}$ and one with \emph{no personalization} $\hat \tau^{U, NP}_{t_1}$ respectively, where $t_0$ is a fixed time by which substantial user learning is expected, and $t_1 = t_0+t_m$ with $t_m$ being the length of a measurement. The system has personalization during $[0,t_1)$. Although the \emph{with personalization} measurement is taken first in time ($t_0 < t_1$), it is adjusted for the lower user learning expected based on historical learning curves, and we denote the final estimate by $\tilde{\tau}^{U, WP}_{t_0}$. The \emph{no personalization} user learning measurement $\hat \tau^{U, NP}_{t_1}$ is otherwise identical to the measurement with personalization, except that a user's past actions and treatment history do not impact the item that the system recommends. 
In other words, $p(\bm X_i^{<t})$ is the same for all $i$ (during the measurement).

As stated in Assumption~\ref{sep}, we assume that the underlying user learning effect is the same regardless of whether the measurement is taken with personalization present or not. In practice, this assumption is thought to be reasonable in this context for two reasons: (i) the personalization setting only differs during the final measurement periods, $[t_0, t_1)$ and $[t_1, t_1+t_m)$, and the measurement duration $t_m$ is typically much shorter than the prior learning period ($t_0$ or $t_1$); and (ii) even when we used a variety of personalization settings during the learning period $[0,t_1)$, we obtained similar $\hat \tau^U_{t_1, NP}$ readings. 

We summarize our findings across these multiple studies in Figure \ref{fig:example_real_world_data} . Each point in the figure corresponds to a different study. On the $y$-axis, we plot the difference in the estimated user learning effects when personalization is turned ``on'', then ``off'', i.e. $\tilde{\tau}^{U, WP}_{t_0} - \hat \tau^{U, NP}_{t_1}\,$, capturing the bias due to personalization. On the $x$-axis, we report a proprietary measure of the difference---or imbalance---in personalization states $p(\bm X_i^{<t})$ between the two cohorts of users being compared, i.e. the $CDT_t$ and $CT$ arms for a CCD experiment. The plotted line is a weighted least squares fit to the data. The intervals represent $95\%$ confidence intervals capturing the uncertainty of each measurement. From Figure~\ref{fig:example_real_world_data}, we confirm that personalization can lead to bias in standard CCD experiments for a real-world large-scale system. Furthermore, we observe that, as the imbalance in personalization between the control and treatment groups increases, the bias of the user learning measurement also increases. Finally, the sign of the personalization discrepancy and user learning bias coincide.

\section{Simulation}
\label{sec:simulation} 

\subsection{Estimator Performance in a Movie Recommendation System}

\begin{figure}
     \begin{subfigure}[b]{0.49\textwidth}
         \centering
         \includegraphics[width=\textwidth]{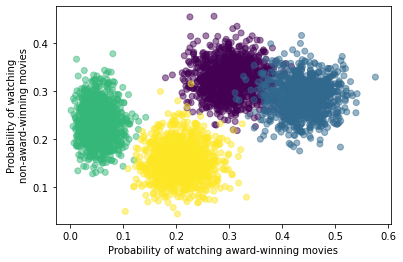}
         \caption{}
         \label{fig:preference}
     \end{subfigure}
     \begin{subfigure}[b]{0.49\textwidth}
         \centering
         \includegraphics[width=\textwidth]{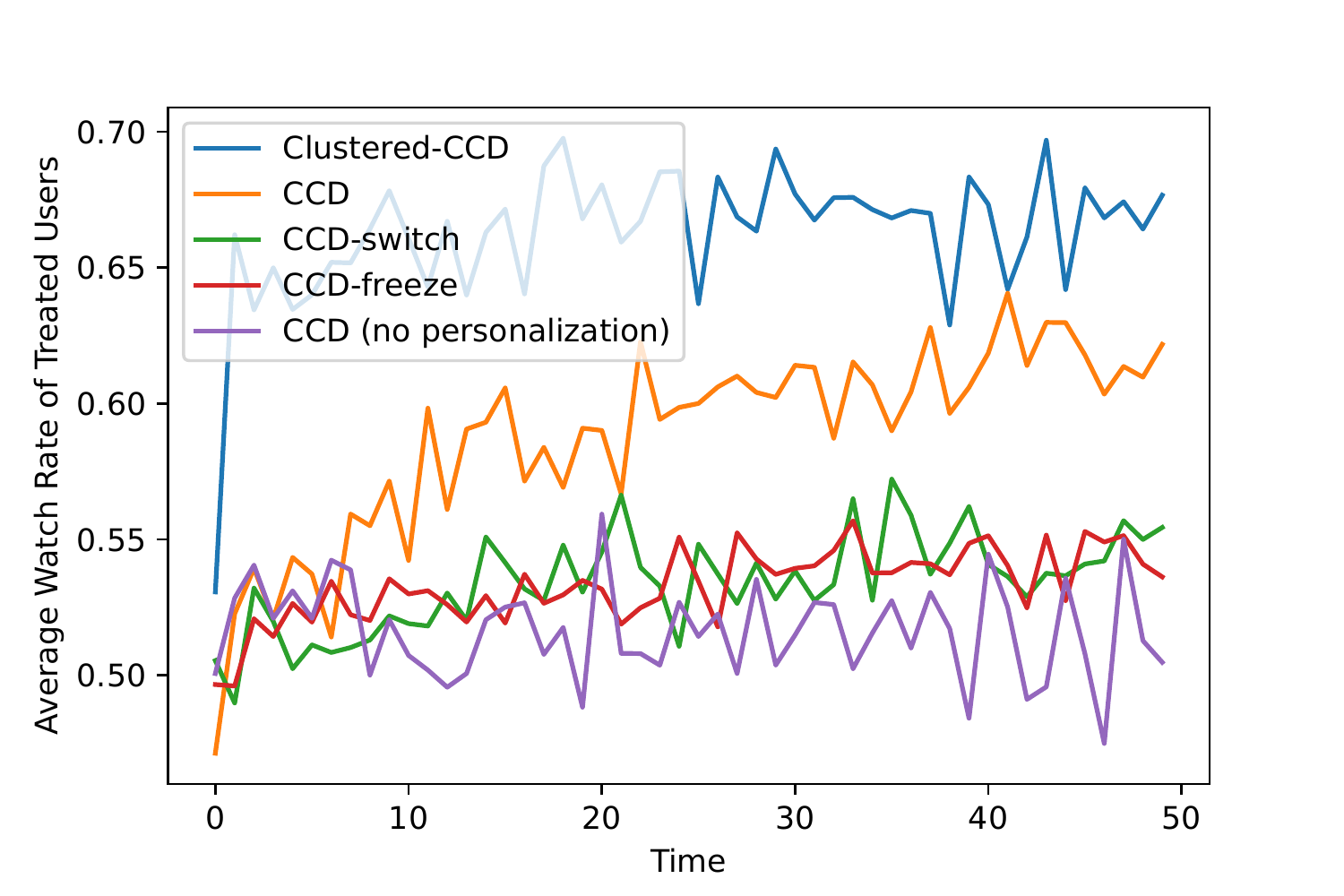}
         \caption{}
         \label{fig:p13ngoodforoutcomes}
     \end{subfigure}
     \caption{(a) User preferences for standard vs. award-winning movies. Each point is a user, with color indicating user cluster. (b) Personalization improves outcomes over time.}
\end{figure}

We now use a simulation of the movie recommendation system experiment discussed in Example \ref{ex:movies} to empirically compare the performance of the user learning estimators discussed in earlier sections, namely: $\hat \tau^{WP,U}_t$ (standard CCD with personalization), $\hat \tau^{NP,U}_t$ (standard CCD without personalization), $\hat \tau^{L,U}_t$ (Clustered-CCD), $\hat \tau^{DS, U}_t$ (CCD-Switch), and $\hat \tau^{CCF, U}_t$ (CCD-Freeze). The simulation has $4,000$ users and there are two categories of movie: award-winning movies ($\textsc{aw}$) and standard movies ($\textsc{s}$). User preferences are sampled according to the following distribution, which exhibits a natural clustered structure of $4$ evenly-sized clusters.
\begin{equation*}
    u_c \sim \mathcal{U}([0.05,0.5]^2) \text{\quad and \quad }
    u_{i,m} \sim \beta\left(u_{c(i), m},~ \frac{100}{1 - u_{c(i),m}}\right)\,
\end{equation*}
where $u_c\in[0.05, 0.5]^2$ is the center of cluster c, $u_{i,m}\in[0,1]$ denotes user $i$'s constant preference for movies of category $m \in \{\textsc{aw}, \textsc{s}\}$, $u_{c(i),m}\in[0,1]^2$ is the m-coordinate of individual i's cluster $c(i)$, $\mathcal{U}$ is the uniform distribution, and $\beta$ is a beta distribution parameterized by its mean and the sum of its two shape parameters. After the cluster centers $u_c$ are drawn, they are shifted so that the average watch rate for each movie type, across all users, is $0.25$. These user preferences are shown in Figure~\ref{fig:preference}, where each point represents an individual, color-indexed by its cluster. The x-axis (resp. y-axis) value of each point represents the user's preference for award-winning (resp. standard) movies.

In the simulation study, we consider a setting where there are $50$ time periods and each user is recommended a single movie per time period. When a user is recommended a movie, they watch it (or not) with some fixed probability, which depends on the movie category, the user's individual preferences, and whether the movie is annotated---the treatment for award-winning movies. The direct effect of the treatment is to increase the probability of a user watching an award-winning movie ($\textsc{aw}$) by a fixed additive constant of $0.5$. Since a standard movie can receive no annotation, a user will see no change in his or her probability to watch a standard movie ($\textsc{s}$) when treated. Formally, the probability $p_{i,t}(m)$ that user i watches movie of category $m \in \{\textsc{aw}, \textsc{s}\}$ at time period $t$ is given by:
\begin{equation}\label{eq:treatment_effect_sim}
   p_{i,t}(m = \textsc{s}) = u_{i,m} \quad\text{and}\quad p_{i,t}(m = \textsc{aw}) = \min(1, u_{i,m} + 0.5 \cdot z_i^t)\,,
\end{equation}
where $z_i^t$ denotes whether user $i$ is treated at time period $t$. There is no direct effect on standard movies. For simplicity, there is also no user learning. 

We allow the movie recommender system to be personalized.  Specifically, there is a 2-tuple user feature $f_{i,t}$ that captures how many times a user $i$ has watched each movie category during all periods up to time $t$. 
The user features are key because, in our simulation, the system chooses which category of movie to recommend to a user based on the probabilities corresponding to the user's feature. Specifically, we choose Thompson Sampling \cite{thompson1933likelihood} with a $\beta(1,1)$ prior as a reasonable heuristic a recommender system might implement to optimize the explore-exploit tradeoff.
We begin all simulations by running the system for 10 periods under the control condition, to gather personalized features for each user. These preliminary features are necessary for the CCD-Freeze method.

User features evolve over time in different ways for each method. For the CCD with no personalization method, user features are fixed at their initial values throughout the entire simulation. For the CCD with personalization method, user features are updated after each time period based on each individual user's actions. For CCD-Switch, the features are similarly updated, except for individuals in the Switch-Cohort. For an individual $i$ in that cohort,  the recommender uses the user feature of the matched individual $j(i)$ in the $CDT_t$ cohort. Users are matched at random from their own cluster, which is a good approximation to an optimal matching algorithm when preferences are naturally clustered, as is the case in our setting. For CCD-Freeze, user features are again updated each time a user watches a movie, except for the features corresponding to users in the Frozen-Cohort which remain fixed at their pre-experiment values throughout the experiment. Finally, for Clustered-CCDs, the shared user feature vector of a cluster is updated every time a user in the cluster watches (or not) a movie. As a result, except for the CCD with no personalization, there is a personalization response to treatment in each experiment design which impacts the movie category the system recommends to users who were exposed to treatment in previous time periods.

In Figure~\ref{fig:p13ngoodforoutcomes}, we demonstrate the benefits of using personalization by plotting the total watch rate of treated users over time. In the first time period, all experiments have similar outcomes, since no treated actions have been recorded by the system. All methods---except for CCD without personalization---see outcomes improve over time as users reveal their preferences to the system. Clustered-CCDs learn preferences very quickly since user preferences are naturally clustered and each user's action is recorded for the entire cluster in each time period. CCD-Freeze and CCD-Switch show slower improvement in watch rate over time than the CCD experiment. The CF and CS cohorts receive recommendations according to a control personalization state even though they receive treatment, which means those cohorts are recommended standard movies more often than they should be, decreasing average watch rates.

\begin{figure}
    \centering
    \includegraphics[clip, trim=4.5cm 0cm 5cm 0cm, width=\textwidth]{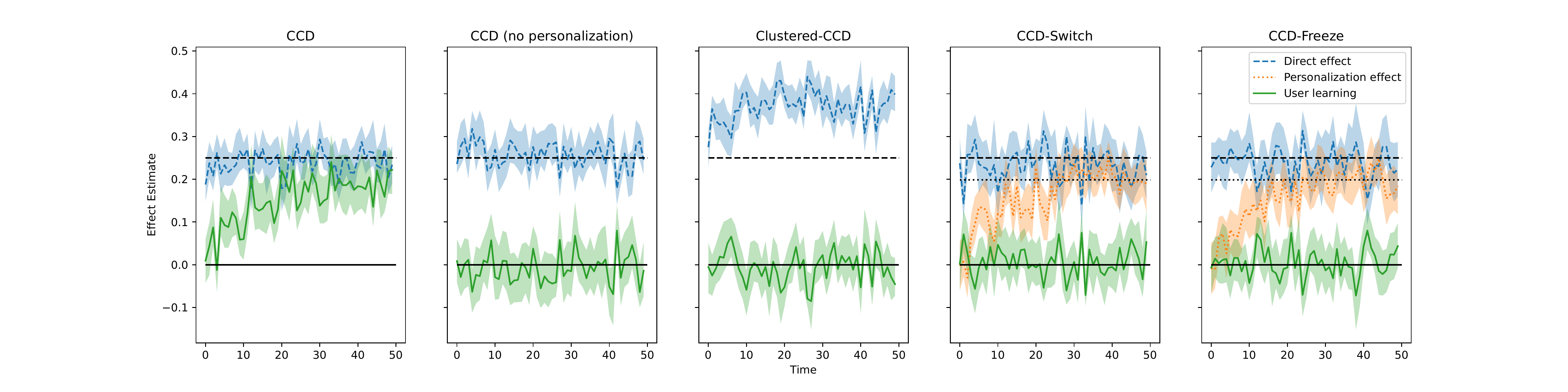}
    \caption{Results for all of the designs. Black lines are the true values (personalization effect is asymptotic).}
    \label{fig:movie_sim}
\end{figure}

In~Figure \ref{fig:movie_sim}, we examine the ability of each experiment design to recover the true estimated direct and user learning effects.  For the CCD-Switch and CCD-Freeze methods only, we also include the estimated personalization effect. Recall that the other approaches do not estimate it. 
The true (asymptotic) personalization effect is computed by comparing two long-term simulations: one with all users in treatment, and another with all users in control except for the final time period, in which they are switched to treatment. Since there is no user learning in our simulation, the difference in average watch rate on the final day captures the personalization effect.
The direct effect is equal to the average incremental watch rate across standard and award-winning movies under the control personalization. Under control, users are on average recommended award-winning and standard movies at equal rates, so $\tau_t^S=0.5 / 2 = 0.25\%$. There is no user learning in this experiment; see Section \ref{sec:extrapolation_experiment} for an experiment that includes user learning.

Each of the CCD-freeze, CCD-switch, and Clustered-CCD methods provide an unbiased estimate of the user learning effect. CCD with no personalization, arguably the simplest design to implement, also leads to unbiased estimates of the user learning and direct effect, but worse outcomes overall as shown in Figure~\ref{fig:p13ngoodforoutcomes}. The simple CCD with personalization leads to a positively-biased estimate of the user learning effect. This occurs because the system recommends more award-winning movies to users in the $CT$ arm  than in the $CDT$ arm, since $CT$ users---treated in the past---watched these at a higher rate. This results in a higher watch rate in the $CT$ arm than in the $CDT$ arm due to differences in the movies being recommended at time $t$. Since the CCD with personalization design does not estimate the personalization effect, it is unable to attribute the increase in watch rate to personalization. All other methods broadly recover the correct zero user learning effect.

Figure~\ref{fig:movie_sim} also shows that all the methods except the Clustered-CCD recover the correct value of the direct effect. The reason for the bias of the Clustered-CCD method is that in this simulation the nature of the treatment breaks the additive separability assumption introduced in Assumption \ref{sep}. Specifically, the simulation violates the separability of personalization and direct effects: the direct effect is higher when users are mostly shown award-winning movies, and lower when users are mostly shown non-award-winning movies. This leads to an over-estimate of the direct effect because, within each cluster, {\it all} users are shown more award-winning movies since some of the users in the clusters are treated and all share the same user features.

Finally, both CCD-Switch and CCD-Freeze estimates of the personalization effects converge to their asymptotic values. The personalization effect increases over time as the system learns the preferences of each user, eventually stabilizing as system estimates of user behavior converge to the users' true behavior. 

\subsection{Extrapolation of Estimated Effects}\label{sec:extrapolation_experiment}

\begin{figure}
 \begin{subfigure}[b]{0.49\textwidth}
         \centering
         \includegraphics[width=\textwidth]{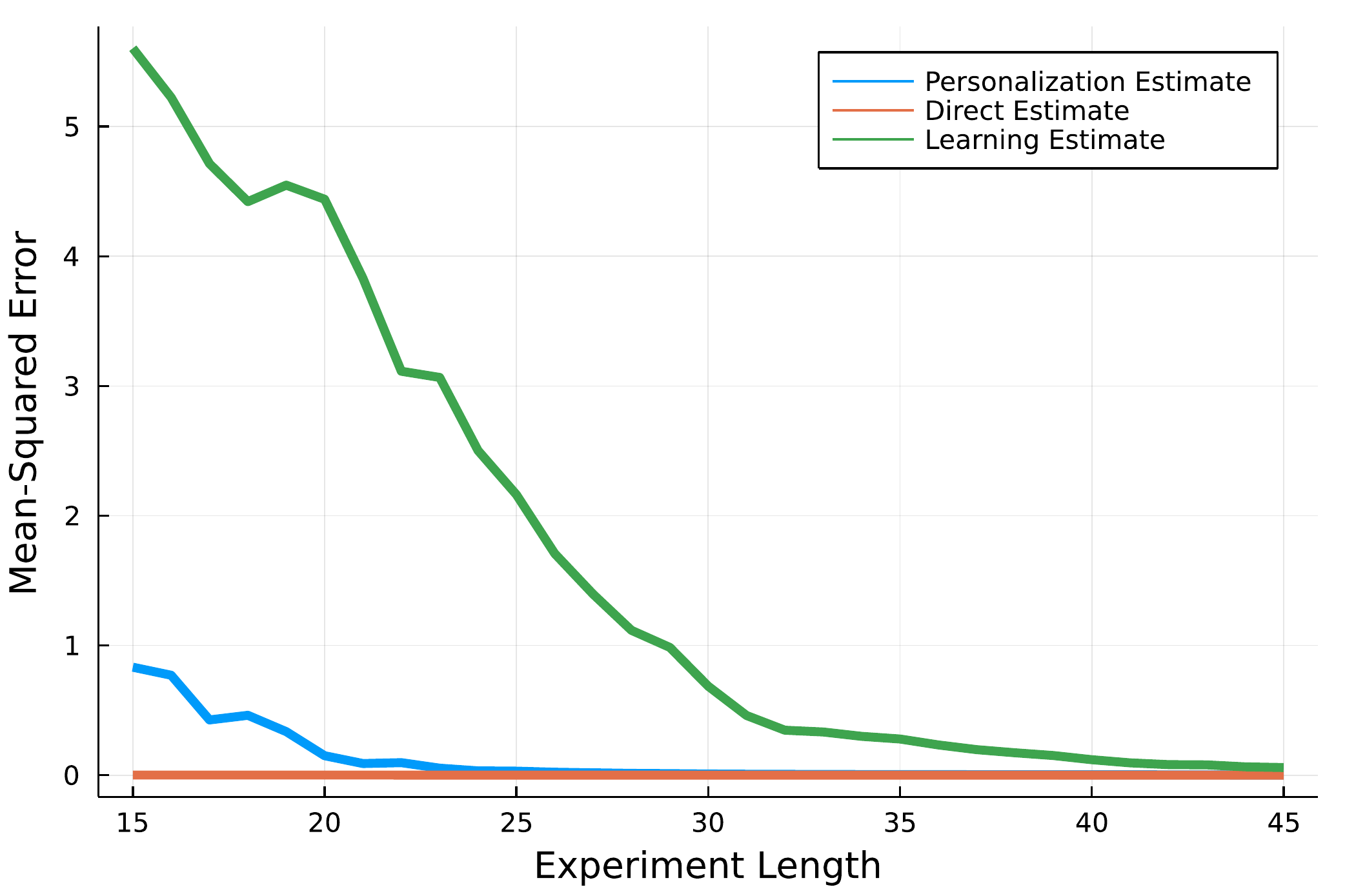}
         \caption{}
         \label{fig:mse}
     \end{subfigure}
     \hfill
     \begin{subfigure}[b]{0.49\textwidth}
         \centering
         \includegraphics[width=\textwidth]{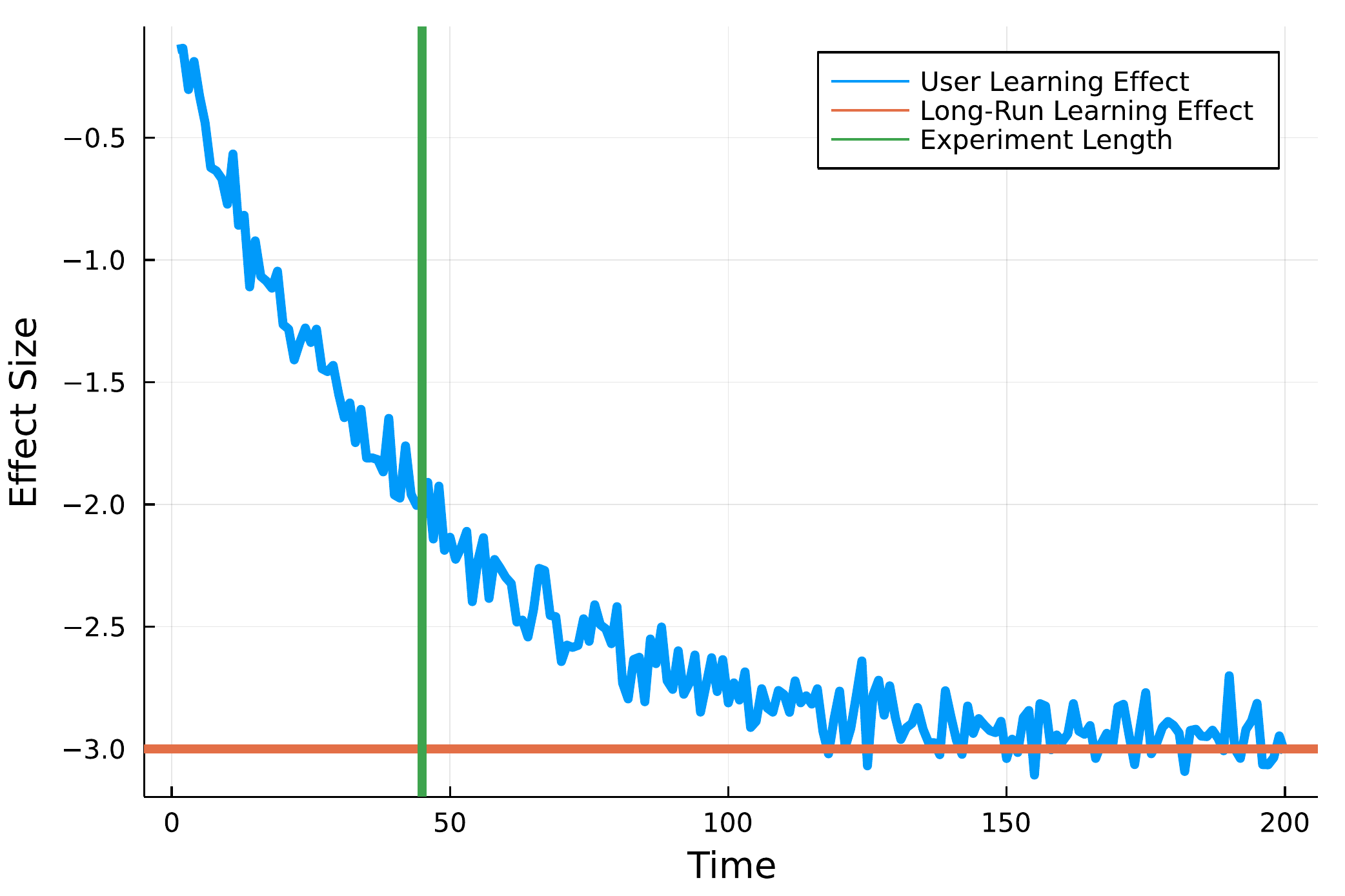}
         \caption{}
         \label{fig:ulearn}
     \end{subfigure}
     \caption{(a) The mean-squared error for the estimators of the long-run effects $\beta^U_0$, $\beta^P_0$ and $\beta^S_0$ converges to 0 for each effect as the horizon of the experiment grows. Results averaged over 300 samples of the data-generating process. (b) The user learning effect converges to its long run value $\beta^U_0$ slowly}
\end{figure} 

The experimental designs in Section \ref{sec:design} can be used to estimate $\tau^P_t$, $\tau^U_t$ and $\tau^S_t$ over time. In this simulation, we show how we can use noisy observations of each of these causal effects over time to estimate a parametric model that can be used to extrapolate to long time horizons. We make the following functional form assumptions, which follow Assumption \ref{as:par}. 
\begin{equation} 
 \tau^U_t = \beta_0^U ( 1 -e^{-\beta_1^U t}) + e^U_t, \quad
  \tau^P_t = \beta_0^P ( 1 -e^{-\beta_1^P t}) + e^P_t, \quad
 \tau^S_t = \beta_0^S + e^S_t.
\end{equation}

We generate data based on this data generating process for $t=1, \ldots, T$ using the parameterization $\beta_0^S = 5.0$, $\beta_0^P = 1.0$, $\beta_0^U = -3.0$, $\beta_1^U = 0.025$ and $\beta_1^P = 0.075$. The long-run direct effect and personalization effects are positive, but are partially offset by the long-run user learning effect. The rate of convergence of the user learning effect is slower than that of the personalization effect. The noise variables $e^q_t$ are drawn from a standard normal distribution with standard deviation 0.1 for each of $q \in \{U, S, P \}$. We vary $T$ from 5 to 45, where $T$ can be thought of as the length of the experiment in days, and examine the mean-squared error of the estimation of $\beta^0_U$, $\beta_0^P$, and $\beta^0_S$ using a non-linear least-squares estimation approach. The results are in Figure \ref{fig:mse}, which is a plot of the mean-squared error, averaged over 100 simulations, of the least-squares estimate of $\beta^U_0$, $\beta^S$ and $\beta^P_0$. These parameters represent the asymptotic user learning, personalization effect, and  direct effect. Since the direct effect is not dynamic, we have very low mean-squared error, even with a short experiment. When only a short time series of the user learning effect is available, however, the MSE is much larger. The user learning effect converges more slowly than the personalization effect, and it requires a larger experiment horizon for MSE to decrease. With a 40-day experiment, however, the estimates of all long-term treatment effects have low error, even though the user learning effect is not yet close to its long-run value, see Figure \ref{fig:ulearn}.

\section{Discussion}\label{sec:discussion}
Interventions in online systems often have a long-term impact that differs from their immediate impact, due to mediated effects that occur slowly through user learning or personalization algorithms. In systems where more complex experiment designs are possible, dynamic mediated effects can be estimated separately, and extrapolated using a parametric model in settings where learning is slow. In future work, it would be interesting to useful to characterize the optimal length of the dynamic experiments for the estimation of user learning when collecting data over longer horizons is costly to the platform.

\bibliographystyle{ACM-Reference-Format}
\bibliography{project.bib} 

\appendix 

\section{Proofs} 

\subsection*{Proof of Proposition \ref{prop:ab}}

\begin{align*} \hat \tau_t &= \frac{1}{n} \sum \limits_{i=1}^n \frac{\mathbbm{1}(G_{it} = T)}{ Pr(G_{it} = T)} Y_{it} -\frac{1}{n} \sum \limits_{i=1}^n \frac{\mathbbm{1}(G_{it} = C)}{ Pr(G_{it} = C)} Y_{it} \\ 
& =\frac{1}{n} \sum \limits_{i=1}^n \frac{\mathbbm{1}(G_{it} = T)}{ Pr(G_{it} = T)} Y_{it}(S_{it}(1, \bm 1^{<t}), U_i(\bm 1^{<t})) -\frac{1}{n} \sum \limits_{i=1}^n \frac{\mathbbm{1}(G_{it} = C)}{ Pr(G_{it} = C)} Y_{it}(S_{it}(0, \bm 0^{<t}), U_i(\bm 0^{<t}))
\end{align*}

$Y_{it}(S_{it}(1, \bm 1^{<t}), U_i(\bm 1^{<t}))$ and $Y_{it}(S_{it}(0, \bm 0^{<t}), U_i(\bm 0^{<t}))$ are fixed potential outcomes. So, taking expectations with respect to the random cohort assignment $G_{it} \in \{C, T\}$:

\begin{align*} 
\mathbb E[ \hat \tau_t] & =\frac{1}{n} \sum \limits_{i=1}^n \left [ \frac{\mathbb E[G_{it} = T)]}{ Pr(G_{it} = T)} Y_{it}(S_{it}(1, \bm 1^{<t}), U_i(\bm 1^{<t})) -  \frac{ \mathbb E[G_{it} = C)]}{ Pr(G_{it} = C)} Y_{it}(S_{it}(0, \bm 0^{<t}), U_i(\bm 0^{<t})) \right ] \\ 
& = \frac{1}{n} \sum \limits_{i=1}^n \left [  Y_{it}(S_{it}(1, \bm 1^{<t}), U_{it}(\bm 1^{<t})) - Y_{it}(S_{it}(0, \bm 0^{<t}), U_{it}( \bm 0^{<t}))\right] \\
& = \tau^{TOTAL}_t. 
\end{align*}

\subsection*{Proof of Proposition \ref{imposs}} 

We start by proving the first part of the proposition. Individuals in $CDT_t$ have $W_{it}=1$ but have not been treated historically, so their outcomes are $Y_{it} = Y_{it}(S_{it}(1, \bm 0^{<t}), U_{it}(\bm 0^{<t}))$. Individuals in $CT$ cohort are currently treated and have been treated historically, so their outcomes are 
$Y_{it} = Y_{it}(S_{it}(1, \bm 1^{<t}), U_{it}(\bm 1^{<t}))$. We plug these into the estimator:

\begin{align*} \hat \tau^{CCD, U}_t & = \frac{1}{n} \sum \limits_{i=1}^n \frac{\mathbbm{1}(G_{it} = CT)}{Pr(G_{it} = CT)} Y_{it} - \frac{1}{n} \sum \limits_{i=1}^n \frac{\mathbbm{1}(G_{it} = CDT_t)}{Pr(G_{it} = CDT_t)  } Y_{it} \\ 
& = \frac{1}{n} \sum \limits_{i=1}^n \frac{\mathbbm{1}(G_{it} = CT)}{Pr(G_{it} = CT)} Y_{it}(S_{it}(1, \bm 1^{<t}), U_{it}(\bm 1^{<t})) - \frac{1}{n} \sum \limits_{i=1}^n \frac{\mathbbm{1}(G_{it} = CDT_t)}{Pr(G_{it} = CDT_t)  } Y_{it}(S_{it}(1, \bm 0^{<t}), U_{it}(\bm 0^{<t}))
\end{align*} 

Then, we take expectations and add and subtract $Y_{it}(S_{it}(1, \bm 0^{<t}), U_{it}(\bm 1^{<t}))$.

\begin{align*} \mathbb E[\hat \tau^{CCD, U}_t] & = \frac{1}{n} \sum \limits_{i=1}^n \left [ \frac{\mathbb E[G_{it} = CT]}{Pr(G_{it} = CT)} Y_{it}(S_{it}(1, \bm 1^{<t}), U_{it}(\bm 1^{<t})) -  \frac{\mathbb E [ G_{it} = CDT_t]}{Pr(G_{it} = CDT_t)  } Y_{it}(S_{it}(1, \bm 0^{<t}), U_{it}(\bm 0^{<t}))\right ] \\ 
& = \frac{1}{n} \sum \limits_{i=1}^n \left [  Y_{it}(S_{it}(1, \bm 1^{<t}), U_{it}(\bm 1^{<t})) - Y_{it}(S_{it}(1, \bm 0^{<t}), U_{it}(\bm 0^{<t}))\right ] \\ 
& = \frac{1}{n} \sum \limits_{i=1}^n \left [  Y_{it}(S_{it}(1, \bm 1^{<t}), U_{it}(\bm 1^{<t})) - Y_{it}(S_{it}(1, \bm 0^{<t}), U_{it}(\bm 1^{<t})  \right] \\   & \qquad + \sum \limits_{i=1}^n \left [Y_{it}(S_{it}(1, \bm 0^{<t}), U_{it}(\bm 1^{<t})  - Y_{it}(S_{it}(1, \bm 0^{<t}), U_{it}(\bm 0^{<t}))\right ] \\ 
& = \tau^P_t + \tau^U_t 
\end{align*} 

Now, we prove the second part of the proposition. Individuals in $CC$ have $W_{it}=0$ and have not been treated historically, so their outcomes are $Y_{it} = Y_{it}(S_{it}(0, \bm 0^{<t}), U_{it}(\bm 0^{<t})) $. Plugging this into the estimator, we have: 
\begin{align*} \hat \tau^{CCD, S} =  \frac{1}{n} \sum \limits_{i=1}^n \frac{\mathbbm{1}(G_{it} = CDT_t)}{Pr(G_{it} = CDT_t)} Y_{it}(S_{it}(1, \bm 0^{<t}) - \frac{1}{n}  \sum \limits_{i=1}^n \frac{ \mathbbm{1}(G_{it} = CC)}{ Pr(G_{it} = CC) } Y_{it}(S_{it}(0, \bm 0^{<t})
\end{align*}

Then taking expectation over the random cohort assignment: 

\begin{align*} 
 \mathbb E[\hat \tau^{CCD, S}_t] & = \frac{1}{n} \sum \limits_{i=1}^n \left [ \frac{\mathbb E[G_{it} = CDT_t]}{Pr(G_{it} = CDT_t)} Y_{it}(S_{it}(1, \bm 0^{<t}), U_{it}(\bm 0^{<t})) -  \frac{\mathbb E [ G_{it} = CC]}{Pr(G_{it} = CC)  } Y_{it}(S_{it}(0, \bm 0^{<t}), U_{it}(\bm 0^{<t}))\right ] \\ 
& = \frac{1}{n} \sum \limits_{i=1}^n \left [  Y_{it}(S_{it}(1, \bm 0^{<t}), U_{it}(\bm 0^{<t})) - Y_{it}(S_{it}(0, \bm 0^{<t}), U_{it}(\bm 0^{<t}))\right ] \\
& = \tau^S_t
\end{align*} 

\subsection*{Proof of Proposition \ref{switch} (Model-Switch)}

The $CS$ cohort receives outcomes $Y_{it} = Y_{it}(s(1, p(\bm X_{j(i)}^{<t})), U_{it}(1, \bm 1^{<t}))$. The $CDT_t$ cohort receives outcomes $Y_{it} = Y_{it}(S_{it}(1, \bm 0^{<t}), U_{it}(\bm 0^{<t}))$. Plugging this into the estimator, 

\begin{align*} 
 \hat \tau^{DS, U}_t = \frac{1}{n}   \sum \limits_{i=1}^n \left [ \frac{ \mathbbm{1}(G_{it} = CS) }{Pr(G_{it} = CS)} Y_{it}(s(1, p(\bm X_{j(i)}^{<t})), U_{it}(1, \bm 1^{<t}))-   \frac{\mathbbm{1}(G_{it} = CDT_t)}{ Pr(G_{it} = CDT_t) } Y_{it}(S_{it}(1, \bm 0^{<t}), U_{it}(\bm 0^{<t})) \right]
\end{align*} 

Then we take expectations over the random cohort assignment, and subtract/add \[Y_{it}(s(1, p(\bm X_{j(i)}^{<t})), U_{it}(1, \bm 1^{<t})) - Y_{it}(S_{it}(1, \bm 0^{<t}), U_{it}(\bm 1^{<t})).\]

\begin{align*} 
\mathbb E[\hat \tau^{DS, U}_t] & = \frac{1}{n} \sum \limits_{i=1}^n \left [ \frac{\mathbb E[G_{it} = CS]}{Pr(G_{it} = CS)}Y_{it}(s(1, p(\bm X_{j(i)}^{<t})), U_{it}(\bm 1^{<t})) -  \frac{\mathbb E [ G_{it} = CDT_t]}{Pr(G_{it} = CDT_t)  } Y_{it}(S_{it}(1, \bm 0^{<t}), U_{it}(\bm 0^{<t}))\right ] \\ 
& = \frac{1}{n} \sum \limits_{i=1}^n \left [  Y_{it}(s(1, p(\bm X_{j(i)}^{<t})), U_{it}(\bm 1^{<t})) - Y_{it}(S_{it}(1, \bm 0^{<t}), U_{it}(\bm 0^{<t}))\right ] \\ 
& = \frac{1}{n} \sum \limits_{i=1}^n \left [  Y_{it}(S_{it}(1, \bm 0^{<t}), U_{it}(\bm 1^{<t})) - Y_{it}(S_{it}(1, \bm 0^{<t}), U_{it}(\bm 0^{<t})) + b^{DS}_t \right]
& = \tau^U_t + b^{DS}_t
\end{align*} 

where $b^{DS}_t = \frac{1}{n} \sum \limits_{i=1}^n Y_{it}(s(1, p(\bm X_{j(i)}^{<t})), U_{it}(1, \bm 1^{<t})) - Y_{it}(S_{it}(1, \bm 0^{<t}), U_{it}(\bm 1^{<t}))$. 

For the second part of the proposition,  individuals in $CT$ cohort are currently treated and have been treated historically, so their outcomes are 
$Y_{it} = Y_{it}(S_{it}(1, \bm 1^{<t}), U_{it}(\bm 1^{<t}))$. Plug the relevant potential outcomes into the estimator: 

\begin{align*} 
\hat \tau^{DS, P}_t = \frac{1}{n}   \sum \limits_{i=1}^n \frac{ \mathbbm{1}(G_{it} = CT) }{Pr(G_{it} = CT)} Y_{it}(S_{it}(1, \bm 1^{<t}), U_{it}(\bm 1^{<t})) -  \frac{1}{n} \sum \limits_{i=1}^n \frac{\mathbbm{1}(G_{it} = CS)}{ Pr(G_{it} = CS) } Y_{it}(s(1, p(\bm X_{j(i)}^{<t})), U_{it}(\bm 1^{<t})) .
\end{align*} 

Taking expectations, and adding/subtracting as before: 

\begin{align*} 
\mathbb E[\hat \tau^{DS, P}_t] & =  \frac{1}{n} \sum \limits_{i=1}^n \left [ \frac{\mathbb E[G_{it} = CT]}{Pr(G_{it} = CT)}Y_{it}(S_{it}(1, \bm 1^{<t}), U_{it}(\bm 1^{<t})) -  \frac{\mathbb E [ G_{it} = CS]}{Pr(G_{it} = CS)  } Y_{it}(s(1, p(\bm X_{j(i)}^{<t})), U_{it}(\bm 1^{<t}))\right ] \\ 
& = \frac{1}{n} \sum \limits_{i=1}^n \left [  Y_{it}(S_{it}(1, \bm 1^{<t}), U_{it}(\bm 1^{<t})) - Y_{it}(s(1, p(\bm X_{j(i)}^{<t})), U_{it}(\bm 1^{<t}))\right ] \\ 
& = \frac{1}{n} \sum \limits_{i=1}^n \left [  Y_{it}(S_{it}(1, \bm 1^{<t}), U_{it}(\bm 1^{<t})) - Y_{it}(S_{it}(1, \bm 0^{<t}), U_{it}(\bm 1^{<t})) - b^{DS}_t \right] \\
& = \tau^P_t - b^{DS}_t \\ 
\end{align*} 

\subsection*{Proof of Proposition \ref{cluster} (Clustering)} 

 In the clustered CCD experiment, the personalization that individual $i$ receives is $p(\bm X_{c(i)}^{<t}))$ and depends on the action history of all individuals in the cluster. $\bm X_{c}^{t}$ depends on the history of user treatments for user $i$ and all other individuals in the cohort. For this proof, we need to make this dependence clear, so we can write the system state under clustering as $s(1, W_{i}^{<t}, \bm W_{-i, c}^{<t})$. Then, the outcomes for the $CDT_t$ cohort are $Y_{it} = Y_{it}(s(1, \bm 0^{<t}, \bm W_{-i, c}^{<t}), U_{it}(\bm 0^{<t}))$ and for the $CT$ cohort are $Y_{it} = Y_{it}(s(1, \bm 1^{<t}, \bm W_{-i, c}^{<t}), U_{it}(\bm 1^{<t}))$. 
 \begin{align*} 
\hat \tau^U_L & = \frac{1}{n} \sum \limits_{i=1}^n \frac{\mathbbm{1}(G_{it} = CT)}{Pr(G_{it} = CT)} Y_{it} - \frac{1}{n} \sum \limits_{i=1}^n \frac{\mathbbm{1}(G_{it} = CDT_t)}{Pr(G_{it} = CDT_t)  } Y_{it} \\ 
& = \frac{1}{n}\sum \limits_{i=1}^n \left [ \frac{\mathbbm{1}(G_{it} = CT)}{Pr(G_{it} = CT)} Y_{it}(s(1, \bm 1^{<t}, \bm W_{-i, c}^{<t}), U_{it}(\bm 1^{<t})) - \frac{\mathbbm{1}(G_{it} = CDT_t)}{Pr(G_{it} = CDT_t)  } Y_{it}(s(1, \bm 0^{<t}, \bm W_{-i, c}^{<t}), U_{it}(\bm 0^{<t})) \right] \\ 
\end{align*} 

Next, we can take expectations over the random cohort assignment. In this case, $Y_{it}(s(1, \bm 0^{<t}, \bm W_{-i, c}^{<t}), U_{it}(\bm 0^{<t}))$ is a random variable since it depends on the cohort assignments of other individuals in the cluster. In the second line, we add and subtract to create a bias term. 

\begin{align*} 
\mathbb E[ \hat \tau^U_L] & = \frac{1}{n}  \sum \limits_{i =1}^n \mathbb E[ Y_{it}(s(1, \bm 1^{<t}, \bm W_{-i, c}^{<t}), U_{it}(\bm 1^{<t}))  - Y_{it}(s(1, \bm 0^{<t}, \bm W_{-i, c}^{<t}), U_{it}(\bm 0^{<t})) ]  
\\ & = \frac{1}{n} \sum \limits_{i=1}^n \mathbb E[ Y_{it}(s(1, \bm 0^{<t}, \bm W_{-i, c}^{<t}), U_{it}(\bm 1^{<t}))  - Y_{it}(s(1, \bm 0^{<t}, \bm W_{-i, c}^{<t}), U_{it}(\bm 0^{<t}))] + b^L_t \\ 
& =^{(1)} \frac{1}{n} \sum \limits_{i=1}^n \mathbb E[ Y_{it}(S_{it}(1, \bm 0^{<t}), U_{it}(\bm 1^{<t}))  - Y_{it}(S_{it}(1, \bm 0^{<t}), U_{it}(\bm 0^{<t}))] + b^L_t.\\
& = \tau^U_t + b^L_t
\end{align*} 

For step (1), we have that the outcomes are evaluated at the same system state $s(1, \bm 0^{<t}, \bm W_{-i, c}^{<t})$ but with different user learning. We can use the additive separability of Assumption \ref{sep} to replace this system state with $S_{it}(1, \bm 0^{<t})$ for each individual under any cohort assignment for the cluster. 

 $b^L_t = \frac{1}{n} \sum \limits_{i=1}^n \mathbb E[Y_{it}(s(1, \bm 1^{<t}, \bm W_{-i, c}^{<t}), U_{it}(\bm 0^{<t})) - Y_{it}(s(1, \bm 0^{<t}, \bm W_{-i, c}^{<t}), U_{it}(\bm 0^{<t}))] $ is driven by the expected difference in the clustered personalized state of an individual who has been historically treated compared to control.  

\section{Clustered-CCD and interference}
\label{sec:interference}

The clustered-CCD experiment design introduces interference. Specifically, the treatment status of a given user affects the personalization state of every other user in its cluster. Under Assumption~\ref{sep}, the user learning effect and direct effect for an individual is the same when measured at any personalization state. Under this assumption, interference does not introduce additional bias in the estimate of the direct and learning effects. See~\citet{savje2021average} for a more detailed discussion of how differences-in-means type estimators are unbiased for direct treatment effects under interference.

\section{Clustering}
\label{app:cluster}

There are multiple ways in which clusters of users could be obtained. We do not aim here to be exhaustive and instead provide a few guiding principles. We first construct a graph of users or place them in a metric space, where the existence of (weighted) edges or the distance in the metric space should be descriptive of the similarity between users. For example, each user's action history can be placed into a metric space, with different coordinates representing averages of actions taken over different time windows. Categorical or demographic information can also be used. Finally, one can go from metric space to a (complete) graph by constructing weighted edges based on the distance between two users in the metric space. 

If metrics spaces are considered, there are many variations of K-means that are both scalable and provide satisfactory results in many instances. If instead a weighted graph is considered, there are many appropriate tools, like balanced partitioning~\cite{aydin2016distributed}. For systems that need to work at scale, we favor hierarchical agglomerative clustering implementations~\cite{murtagh2012algorithms}.

\section{Example of Spurious Positive or Negative Learning} 
\label{app:example}
\subsection*{ Annotating award-winning movies from the in-house studio}

Suppose instead that the streaming service seeks to use the ``award-winning movie'' annotation to improve viewers' opinions of its own in-house film studio. The annotation will be placed on eligible in-house movies but not on externally produced movies. The streaming service once again expects an immediate increase in watch rate on annotated movies (the direct effect). They also hypothesize that there will be a long-term increase in watch rate on in-house movies, as users watch high-quality in-house movies and improve their opinion of the in-house studio. The long-term user learning effect will be measured with a CCD experiment comparing the watch rate on in-house movies in the CT vs CDT cohorts.

\begin{figure}
    \centering
    \includegraphics[clip, trim=0cm 8.5cm 4.5cm 0cm, width=\textwidth]{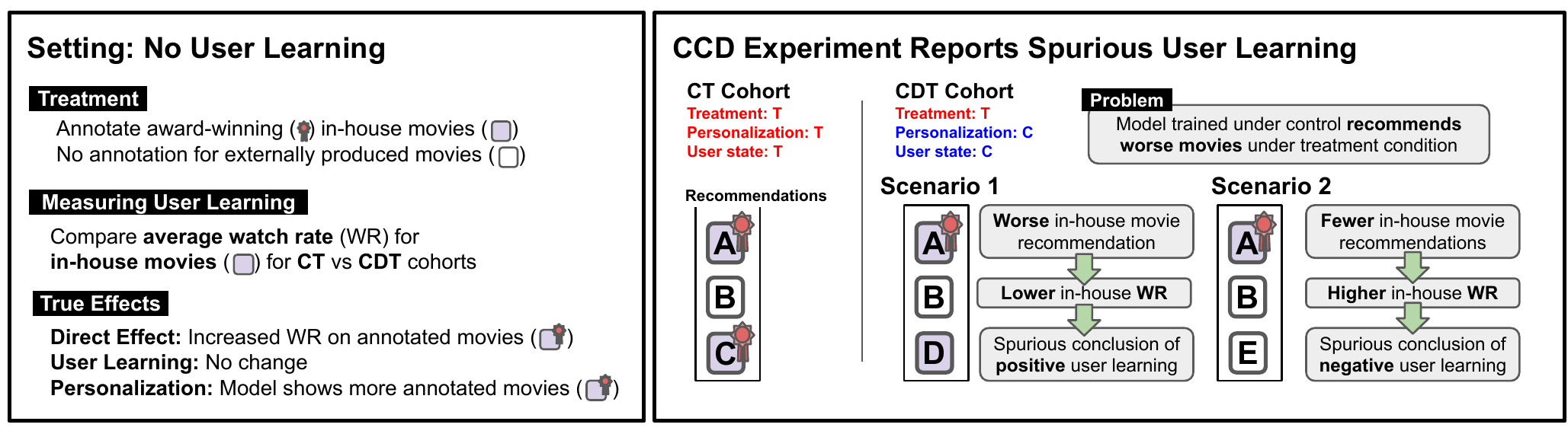}
    \caption{Personalization effects may lead to positive or negative bias when user learning is estimated with a CCD experiment.}
    \label{fig:movie-rec}
\end{figure}

In this example, personalization may cause the CCD experiment to measure either spurious positive or spurious negative user learning. Both types of errors are illustrated in Figure \ref{fig:movie-rec}.

\textbf{Spurious positive learning:} The mechanism for spurious positive learning is similar to that of Example 1. Because there is no user learning, both CT and CDT users have an equal chance of watching award-annotated in-house movies, and this chance is higher than it would be if the movie were not annotated. However, the personalization model knows that CT users have this preference, whereas it is unaware of the preference for CDT users. As a result, CT users get recommendations that better fit their preferences, resulting in a higher rate of watching in-house movies for CT users than for CDT users. The experiment spuriously concludes positive user learning.

\textbf{Spurious negative learning:} The mechanism for spurious negative learning also relies on better recommendations for the CT cohort, but in this case it depends on CDT users being shown fewer in-house movies of higher average quality. Suppose the recommendation algorithm reacted to the treated users' increased affinity for awarded in-house movies by recommending those movies more frequently. The difference between treatment and control recommendations now becomes that the treated users have more awarded in-house movies, and as a result the in-house movies shown to treatment but not control will be of lower quality than the average movie shown to control. This pushes the in-house movie watch rate of treated users down, and could lead to spurious conclusions of negative user learning.

\end{document}